\documentclass{article}
\usepackage{float}
\usepackage{graphicx}
\usepackage{amsmath,amssymb,amsthm}
\usepackage[colorlinks=true,linkcolor=blue,citecolor=blue,urlcolor=blue]{hyperref}

\newcommand{\ii}{\mathrm{i}}

\newtheorem{remark}{Remark}[section]

\title{Stable soap bubble clusters with multiple torus bubbles}
\author{Fabrice Delbary\\\href{mailto:fabrice.delbary@cnrs.fr}{fabrice.delbary@cnrs.fr}}
\date{}

\begin{document}
\maketitle

\section{Abstract}
In the last two centuries and more particularly in the last decades, the geometry of foams has become an important research domain, in mathematics, physics, material sciences and biology \cite{cantat_cohen-addad_elias_graner_hoehler_pitois_rouyer_saint-jalmes_flatman_cox,morgan-book,weaire_hutzler}. Most of the simplest geometrical observations of bubble clusters have long resisted rigorous mathematical proofs. Geometries can even get more complicated if immiscible fluids are considered. Although they have to fulfill Plateau's laws \cite{almgren-taylor,plateau} like soap bubble clusters if the surface tensions are close to unity \cite{morgan-paper}, this is not the case in general \cite{morgan-book}. In 1996, Frederick J. Almgren asked whether there is ``any stable cluster of bubbles in $\mathbb{R}^3$ with some bubble being topologically a torus'' \cite{sullivan-morgan} (problem 1). We propose to answer the latter numerically with simple examples. We build stable soap bubble clusters with a triple torus bubble, a fivefold torus bubble or an elevenfold torus bubble. The construction uses the geometry of a simple immiscible fluids cluster with a torus bubble.
\section{Introduction}\label{intro}
Finding the geometry of a static foam consists in finding an energy minimizer (possibly local) of the bubble cluster for given volumes enclosed by each bubble. The energy is defined as the sum of the areas of the cluster interfaces. For immiscible fluids, each interface area can be weighted with a surface tension. One of the very first question about ``foam'' geometry is whether the smallest area enclosing a given volume is the sphere. This was formulated by Archimedes in the third century B.C.. This is only in the nineteenth century that it was proven to be true by Hermann A. Schwarz in 1884 \cite{schwarz}. The double bubble conjecture was formulated in the nineteenth century and states that the surface of minimal area enclosing two volumes is the standard double bubble, i.e. composed of three spherical caps sharing a circle intersection. The proof dates back from 2002 in \cite{hutchings-morgan-ritore-ros}. The next natural question coming to mind is if larger clusters are also composed of spherical interfaces. In 2022, a proof \cite{milman-neeman} shows that the only bubble clusters of at most $3$ bubbles are the standard clusters, hence proving a conjecture of John M. Sullivan in 1996 \cite{sullivan-morgan} (problem 2). It implies that clusters of at most $3$ bubbles only contain spherical interfaces. The problem for clusters of $4$ and $5$ bubbles remains open. For $6$ bubbles, an example \cite{almgren_sullivan,morgan-book,sullivan} by John M. Sullivan exhibits a cluster containing a saddle shape interface hence shows that clusters of at least $6$ bubbles can contain non spherical interfaces. In fact, numerical simulations tend to show that large clusters are rarely composed of spherical interfaces exclusively and maybe even rarely contain any spherical interface at all. Another interesting question by Frederick J. Almgren in 1996 \cite{sullivan-morgan} (problem 1) is whether there is ``any stable cluster of bubbles in $\mathbb{R}^3$ with some bubble being topologically a torus''\footnote{Note that torus bubbles have long been observed in the framework of dynamics \cite{walters_davidson}.}. We propose to tackle this problem in this document. We start with building a $5$-bubble cluster for immiscible fluids with a torus bubble. This cluster geometry is then used as a basis to build stable soap bubble clusters with a triple, a fivefold or an elevenfold torus bubble. To do so, we duplicate the $5$-bubble cluster and assemble the copies around a tetrahedron, a cube or a dodecahedron.
\section{Torus bubble for immiscible fluids}\label{torim}
\subsection{Surface Evolver configuration}
The first cluster we construct contains three bubbles having an exterior interface, the one in the middle is of genus $1$. Inside the cluster there are two other bubbles. All but two surface tensions are equal to $1$. The surface tension of the interfaces shared by the outer bubbles is $1/4$. All simulations and Hessian matrix eigenvalues computations are performed using Surface Evolver (Kenneth Brakke \cite{brakke}) and double precision. The initial configuration is shown in the following figure.
\begin{figure}[H]
\centering
\includegraphics[scale=0.18]{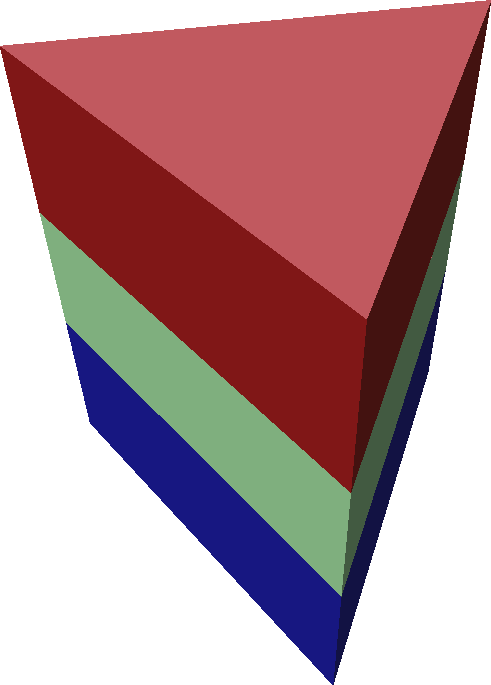}\includegraphics[scale=0.18]{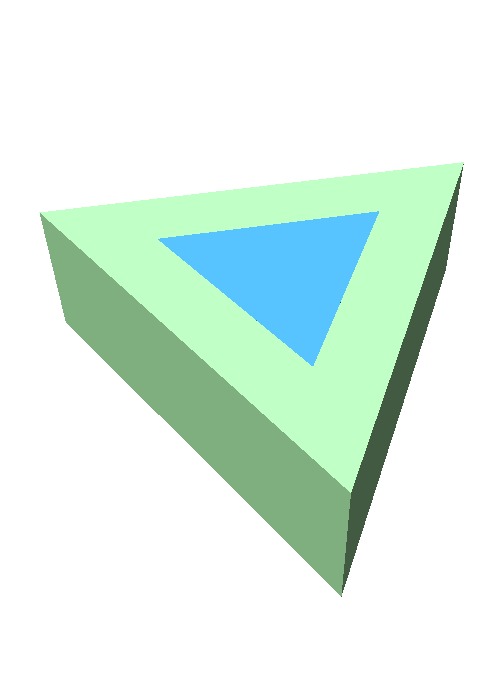}\includegraphics[scale=0.18]{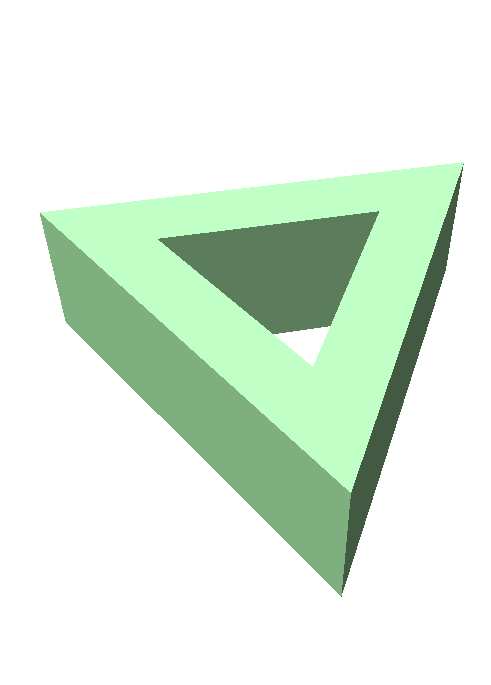}\hspace{-0.6cm}\includegraphics[scale=0.18]{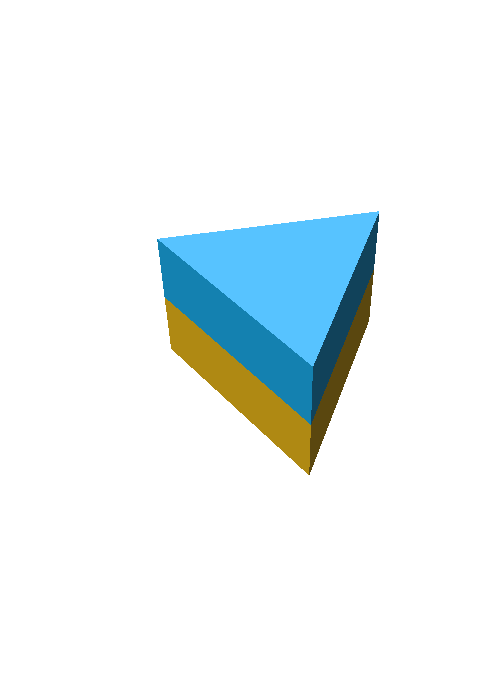}
\caption{From left to right: outer bubbles, torus and inner double bubble, torus bubble, inner double bubble.}
\label{initgen1}
\end{figure}
The cluster is symmetric with respect to the $(0,x,y)$ plane. In this plane, the outer triangles have coordinates $\exp(2\ii k\pi/3)/2, k=0\ldots2$ and the inner ones $\exp(2\ii k\pi/3)/4, k=0\ldots2$. Along the $z$-axis, the height of each inner bubble is $0.15$, the height of the outer middle bubble (of genus $1$) is $0.3$ and the height of the outer upper and lower bubbles is $0.35$. One can illustrate the configuration using a smoothed slice.
\begin{figure}[H]
\centering
\includegraphics[scale=0.35]{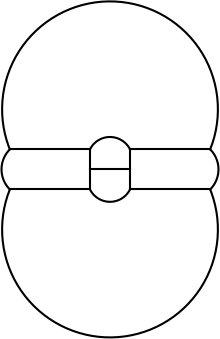}
\caption{$2$D view of the cluster.}
\end{figure}
\subsection{Energy minimizing cluster}\label{genus_01_emc}
After minimization, we get the following cluster configuration (total area $3.54$ and energy $2.98$).
\begin{figure}[H]
\centering
\includegraphics[scale=0.17]{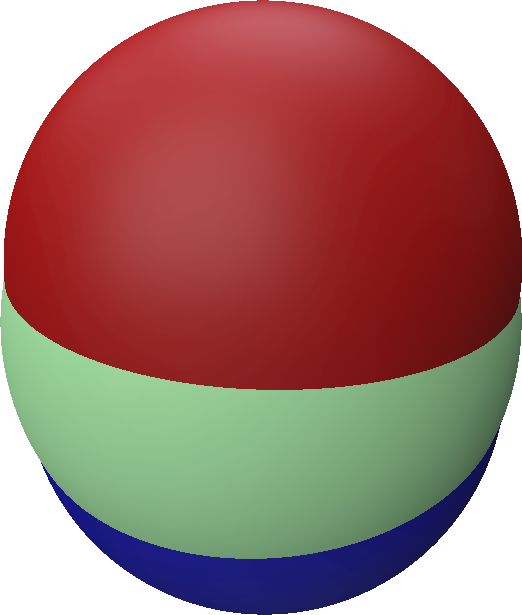}\hspace{0.2cm}\includegraphics[scale=0.17]{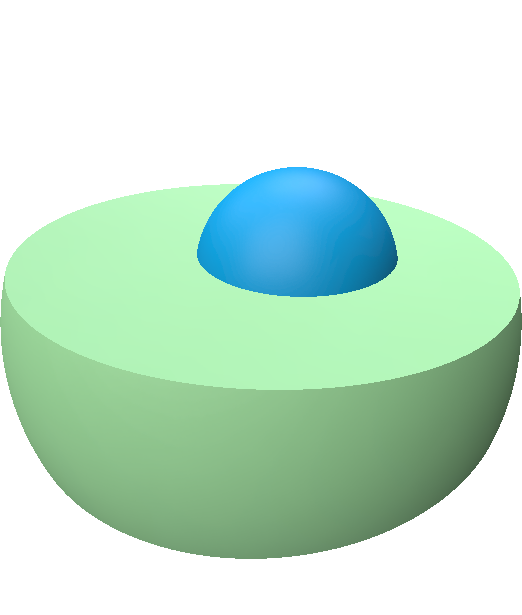}\hspace{0.2cm}\includegraphics[scale=0.17]{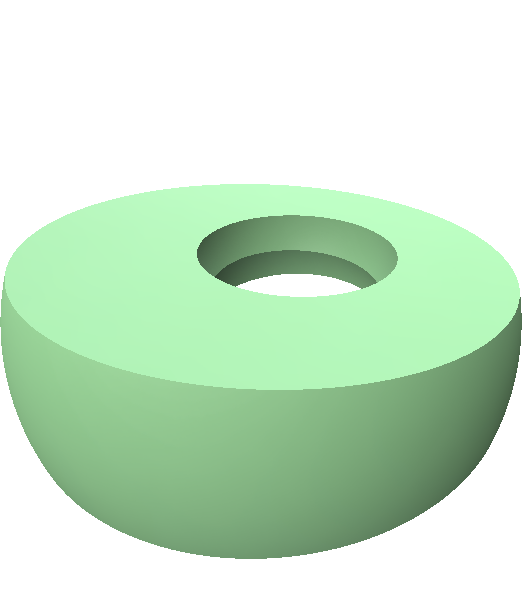}\hspace{-0.9cm}\includegraphics[scale=0.17]{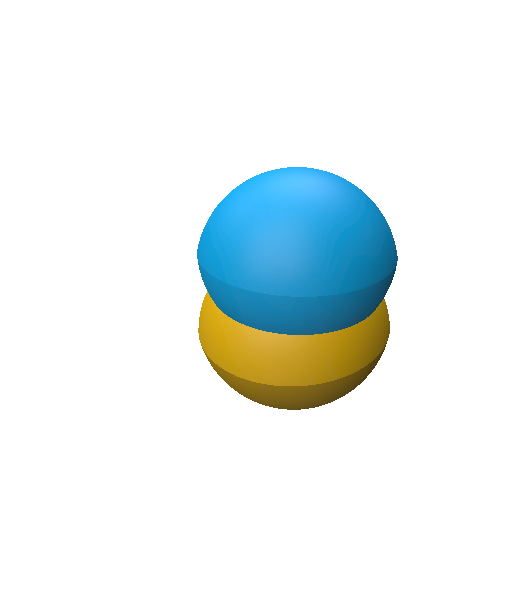}
\label{end_genus_01}
\end{figure}
The cluster which is obtained here is not axisymmetric hence although of genus $1$, the middle bubble is not a toroid. However, two of the symmetry planes seem to be conserved. Moreover, comparing the interfaces to their mean plane or mean sphere indicates that the interfaces may be planar or spherical. Although more numerical tests can be done, no axisymmetric cluster minimizing the energy has been found. When all surface tensions are $1$, clusters with this geometry (axisymmetric or not) do not seem to minimize the energy either. During the minimization process, either the inner double bubble is cut or slides towards the outer interfaces, hence producing non-Plateau borders when reaching them. Different discretizations have been tested. For each one, the simulation is stopped when the energy minimum is reached (positive Hessian and no configuration change for Hessian steps). The discretizations are defined by the longest triangle edge of the mesh. The lengths we consider are $1.6\times10^{-2},8\times10^{-3},4\times10^{-3},2\times10^{-3}$. The double bubble interfaces shared with the torus bubble need a fine discretization, we can check that the ones we chose are relevant. The two coarsest meshes are plotted in the following figures.
\begin{figure}[H]
\centering
\includegraphics[scale=0.1]{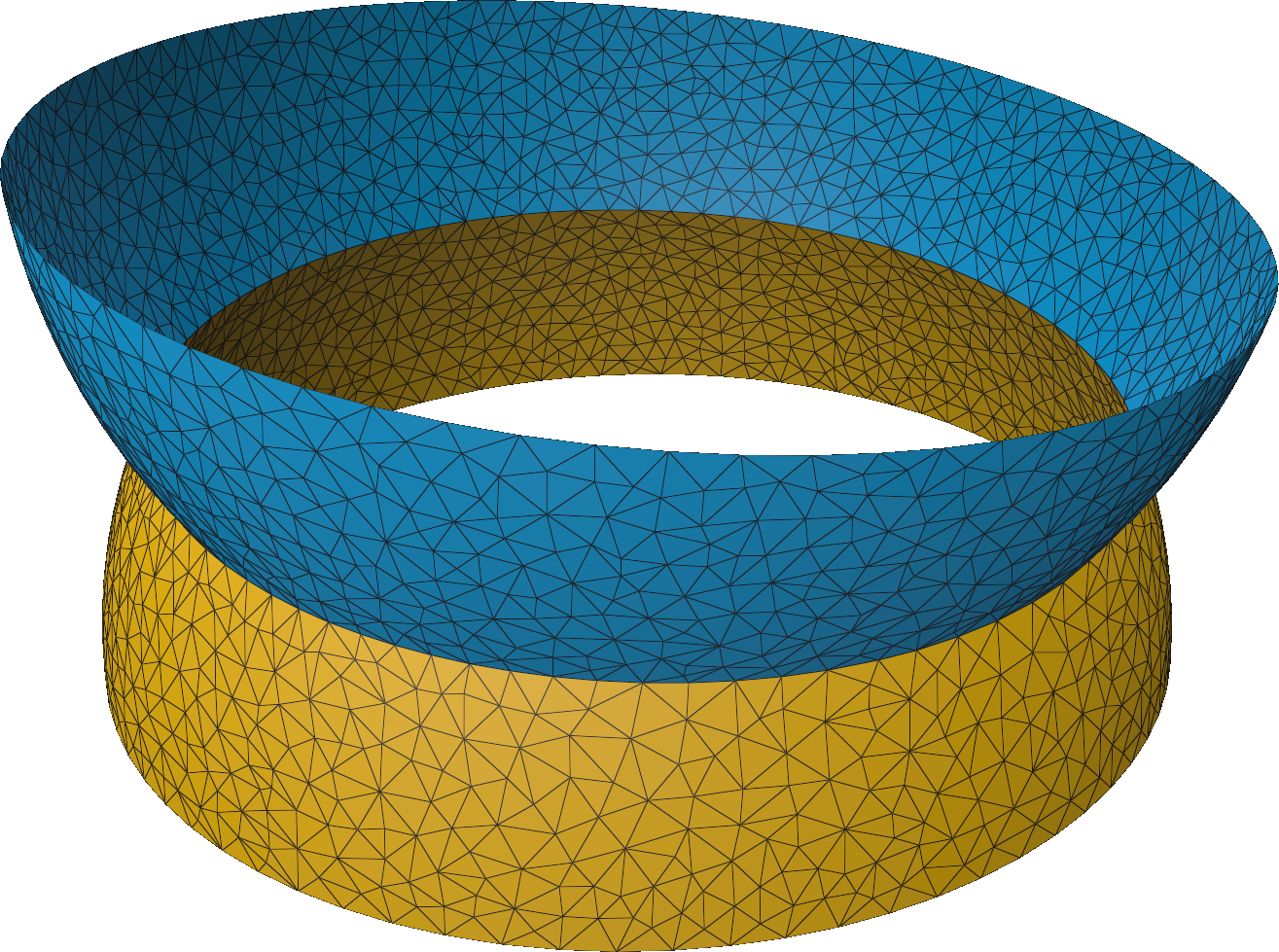}\hspace{2cm}\includegraphics[scale=0.1]{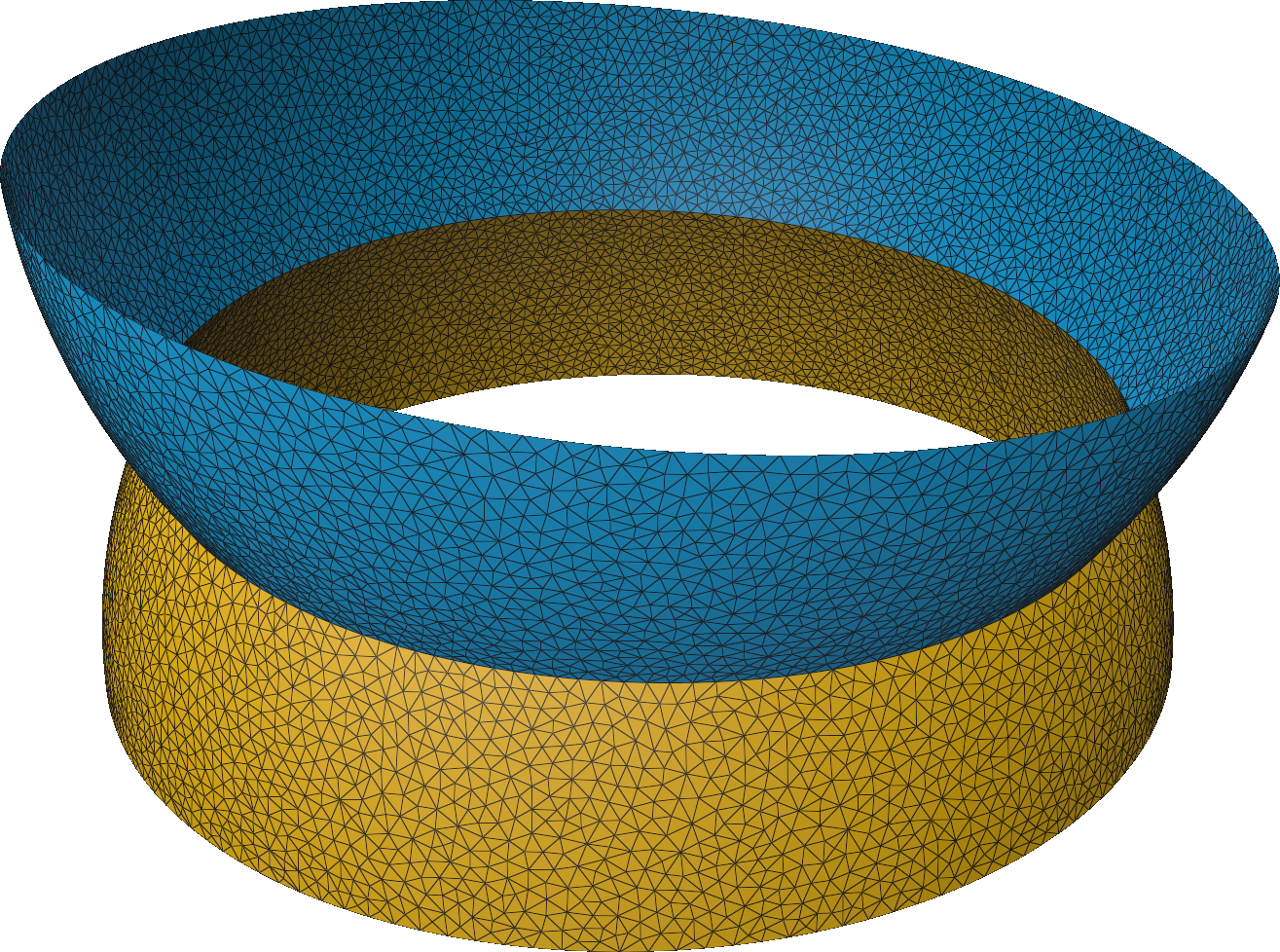}
\caption{Left discretization: $1.6\times10^{-2}$. Right discretization: $8\times10^{-3}$.}
\end{figure}
For each discretization, the Hessian matrix lowest and highest eigenvalues at the minimum are computed.
\begin{table}[H]
\centering
\begin{tabular}{|c|c|c|}
\hline
Discretization & Lowest eigenvalue & Highest eigenvalue\\
\hline
$1.6\times10^{-2}$ & $2.1\times10^{-7}$ & $10$\\
\hline
$8\times10^{-3}$ & $1.9\times10^{-8}$ & $9$\\
\hline
$4\times10^{-3}$ & $2.3\times10^{-9}$ & $10$\\
\hline
$2\times10^{-3}$ & $3.3\times10^{-10}$ & $11$\\
\hline
\end{tabular}
\caption{Hessian matrix eigenvalues for the different discretizations.}
\end{table}
The orders of magnitude of the minimum eigenvalues are what can be expected in comparison to other clusters with similar bubble volumes, interfaces area, surface tensions and discretization. Although the latter are not the only parameters the eigenvalues depend on, it is quite reasonable to assume that the stability of the cluster is not a numerical artifact.
\section{Multiple torus bubbles for soap bubble clusters}
\subsection{Triple torus bubble}
\subsubsection{Surface Evolver configuration}\label{tetconf}
As mentioned in the section \ref{genus_01_emc}, when all surface tensions of the previous cluster are $1$, if the two inner bubbles do not separate, they slide towards the outer interfaces, producing non-Plateau borders when reaching them. Hence, a possible idea to get a stable cluster including the previous geometry would be to ``constrain''  it on the sides. To this end, one can duplicate the cluster around a tetrahedron. More precisely, reshaping the previous cluster as a cone (instead of a right triangular prism), we assemble four of them in the following way.
\begin{figure}[H]
\centering
\includegraphics[scale=0.12]{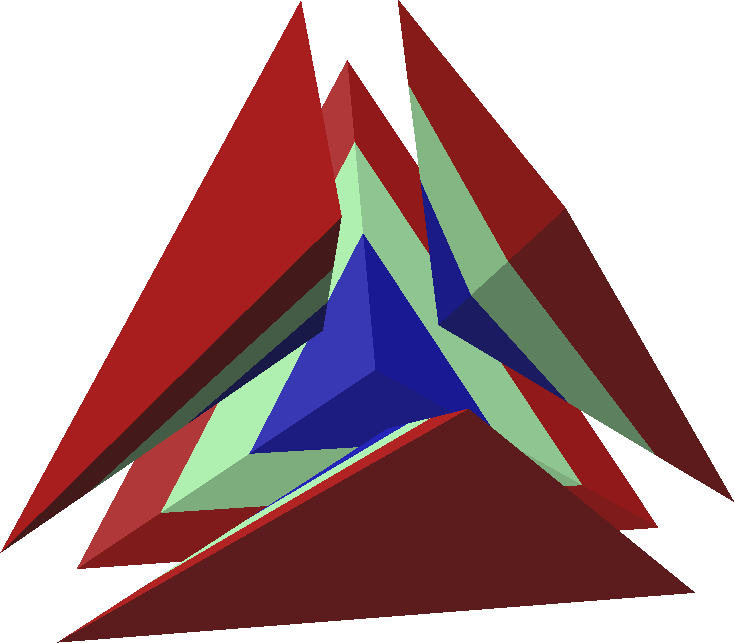}\includegraphics[scale=0.12]{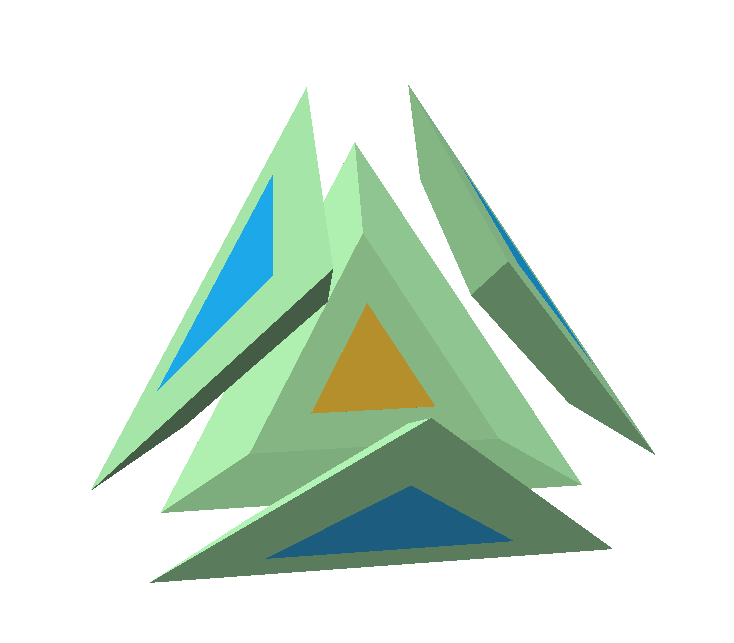}\hspace{-0.5cm}\includegraphics[scale=0.12]{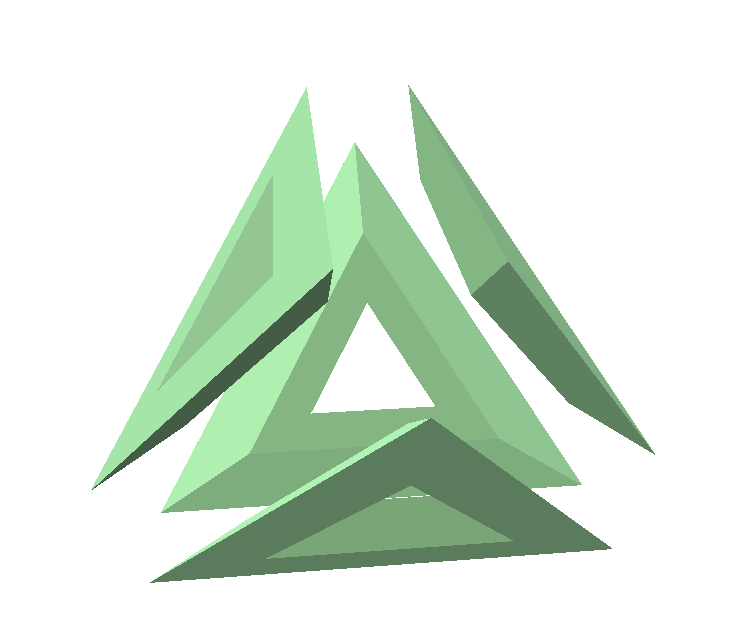}\hspace{-1cm}\includegraphics[scale=0.12]{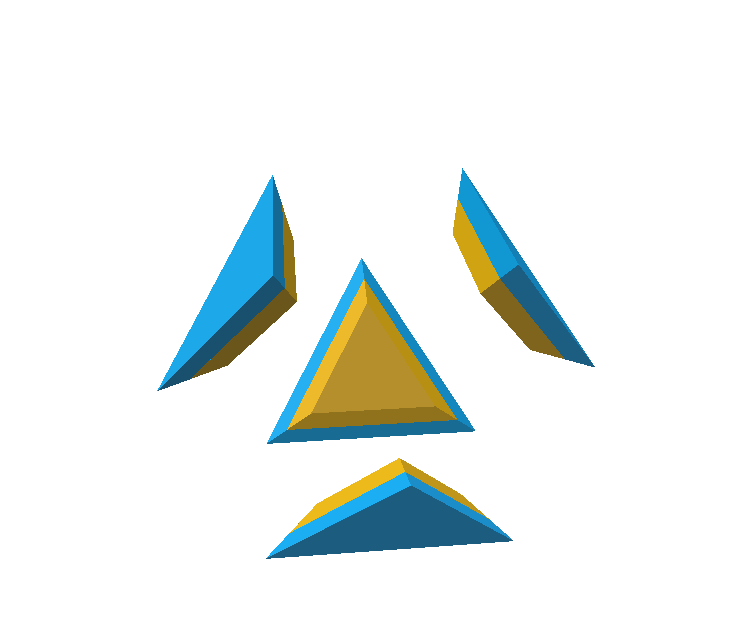}
\caption{From left to right: the four clusters, torus and inner double bubbles of each cluster, torus bubble of each cluster, inner double bubbles of each cluster.}
\label{assemble}
\end{figure}
The four bubbles at the tip of the cones become a unique bubble and the four tori become a unique bubble having the shape of an excavated tetrahedron surface. Hence the whole cluster has $14$ bubbles. The following figures give different views of the cluster by hiding some bubbles. We start with a view of the outer bubbles, then hide one of them, afterwards we also hide the multiple torus bubble and finally we also hide the inner double bubble below the hidden outer bubble.
\begin{figure}[H]
\centering
\includegraphics[scale=0.13]{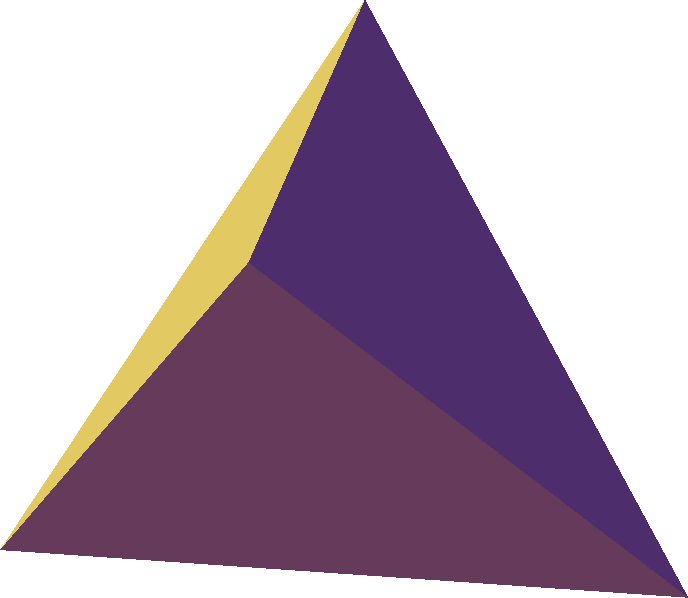}\hspace{-0.7cm}\includegraphics[scale=0.13]{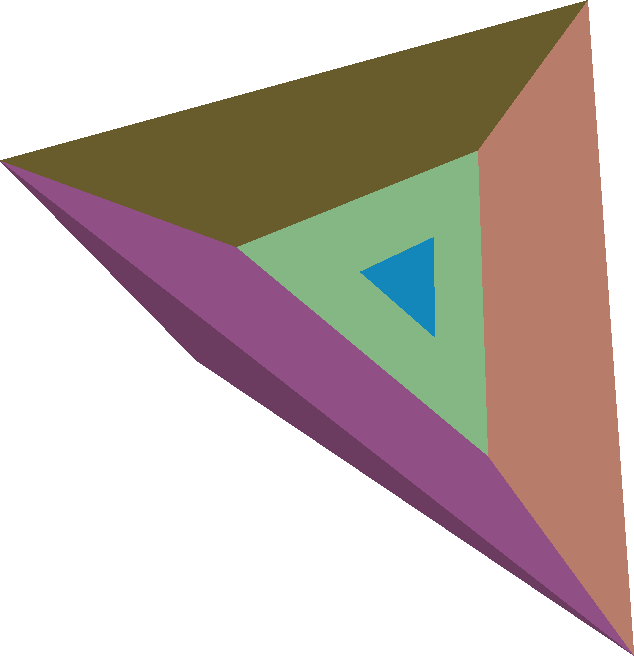}\hspace{0.2cm}\includegraphics[scale=0.13]{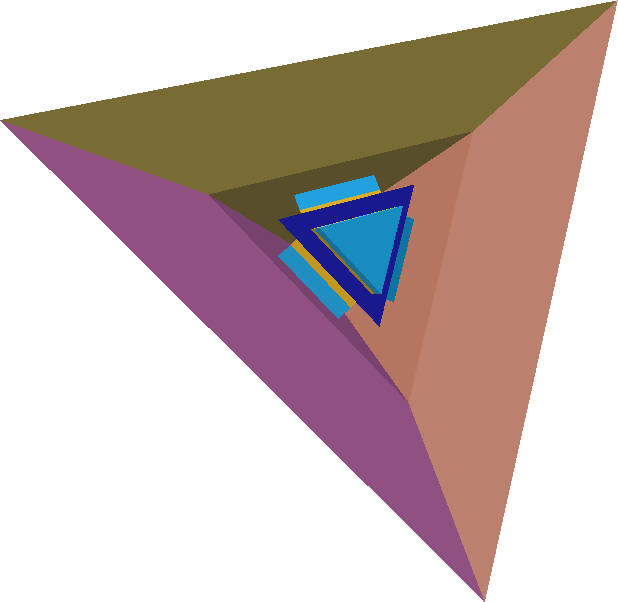}\hspace{0.2cm}\includegraphics[scale=0.13]{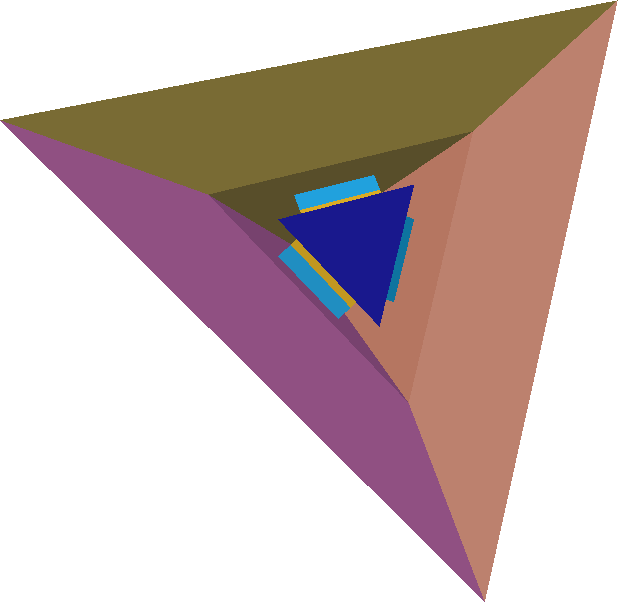}
\caption{Different views of the initial cluster.}
\end{figure}
\begin{figure}[H]
\centering
\includegraphics[scale=0.15]{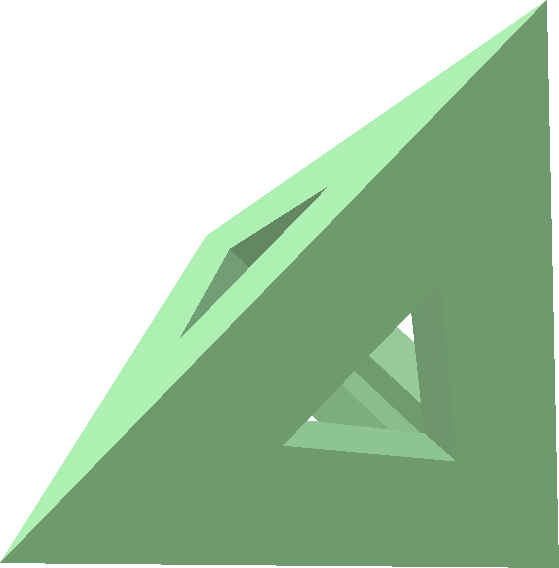}
\caption{Triple torus bubble.}
\label{tetmultor}
\end{figure}
\begin{remark}
Compared to the assembled initial cluster of figure \ref{assemble}, the inner double bubbles are chosen to be right triangular prisms instead of triangular frusta.
\end{remark}
A smoothed $2$D schematic view of the configuration might help to get a better understanding.
\begin{figure}[H]
\centering
\includegraphics[scale=0.35]{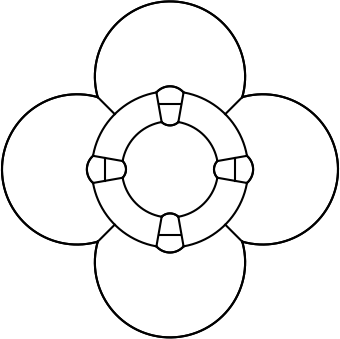}
\caption{Smoothed $2$D schematic view of the cluster.}
\end{figure}
Figuratively speaking, one has a center bubble surrounded by another one. Outer bubbles are then attached to the center one with ``snap fasteners'' (inner double bubbles) which pierce the surrounding bubble.\\
The genus of a compact, closed, connected, orientable topological surface without boundary is defined as the largest number of simple closed curves with no common points that can cut the surface without making it disconnected. Simply put, the genus is the ``number of holes'' in the surface (in the sense of torus-like holes). However, consider now a solid whose boundary has genus $0$ (sphere). Dig a cavity in the solid, with no contact with the exterior, and $n$ holes from outside to the cavity. Then the surface of the new solid has genus $n-1$, i.e. is topologically an $(n-1)$-fold torus. Roughly speaking, one can imagine that one of the holes is stretched until the surface can be placed on a plane, resulting in $n-1$ holes only. This can be illustrated in the following way for an excavated tetrahedron whose surface is homeomorphic to the bubble of figure \ref{tetmultor}.
\begin{figure}[H]
\centering
\includegraphics[scale=0.1]{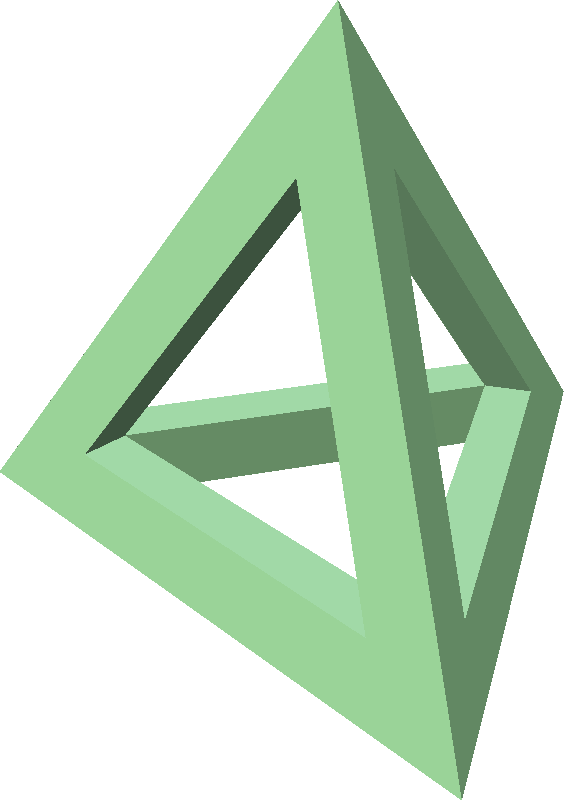}\hspace{0.5cm}\includegraphics[scale=0.08]{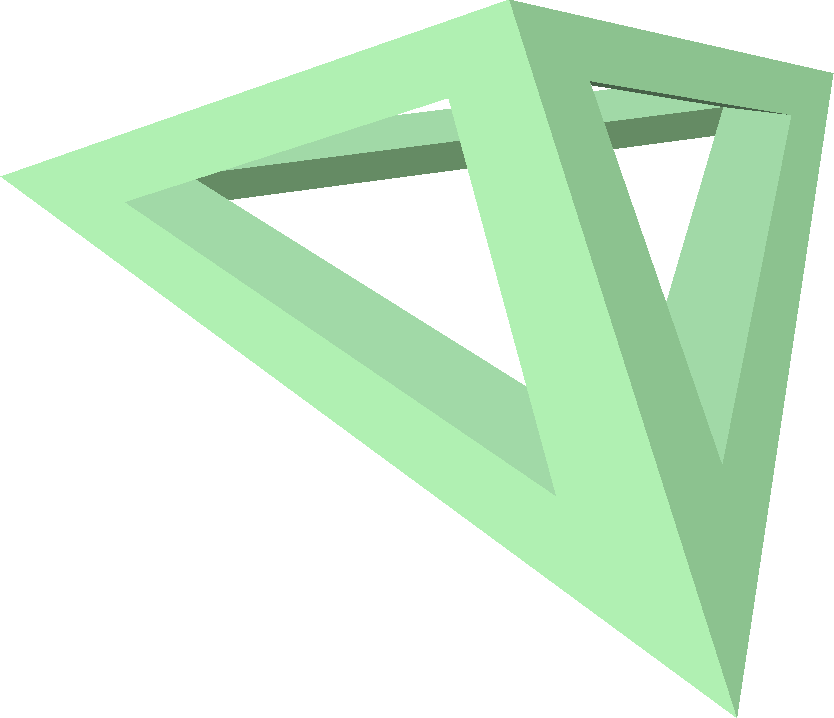}\hspace{0.5cm}\includegraphics[scale=0.1]{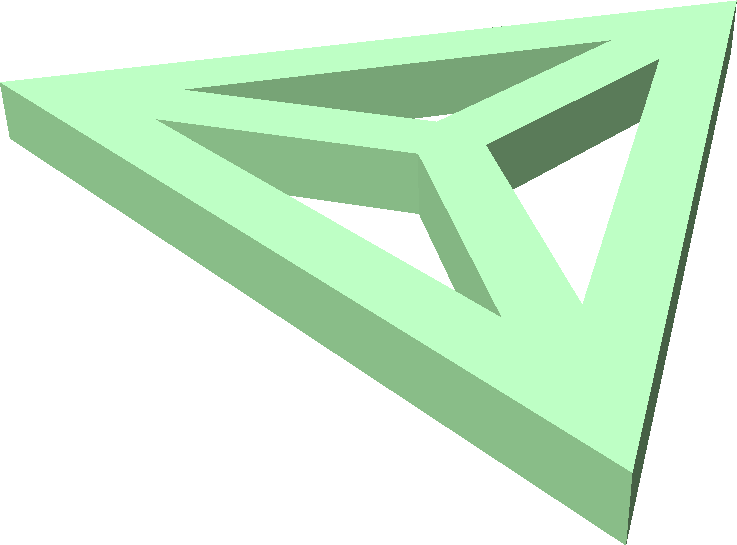}\hspace{0.5cm}\includegraphics[scale=0.14]{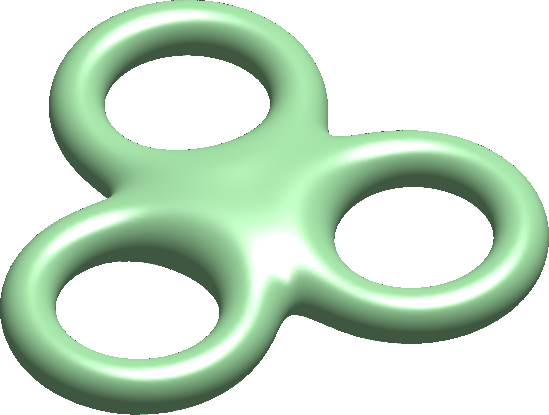}
\caption{Continuous transformation of the excavated tetrahedron surface into a triple torus.}
\label{conttrans}
\end{figure}
Hence the bubble of figure \ref{tetmultor} is topologically a triple torus. One can concisely and more precisely prove this result using the Euler characteristic of the excavated tetrahedron of figure \ref{conttrans}. The Euler characteristic $\chi$ of a polyhedron is defined by
\begin{equation}
\chi=V-E+F,
\end{equation}
Where $V,E,F$ are respectively the number of vertices, edges and faces. If the surface of the polyhedron is a topological surface, the Euler characteristic of the polyhedron is linked to the genus $g$ of its surface by
\begin{equation}
\chi=2(1-g).
\end{equation}
As is, the excavated tetrahedron of figure \ref{conttrans} is not really a polyhedron since the exterior faces are not simple polygons. Hence, we divide them into three isosceles trapezoids in the following way.
\begin{figure}[H]
\centering
\includegraphics[scale=0.12]{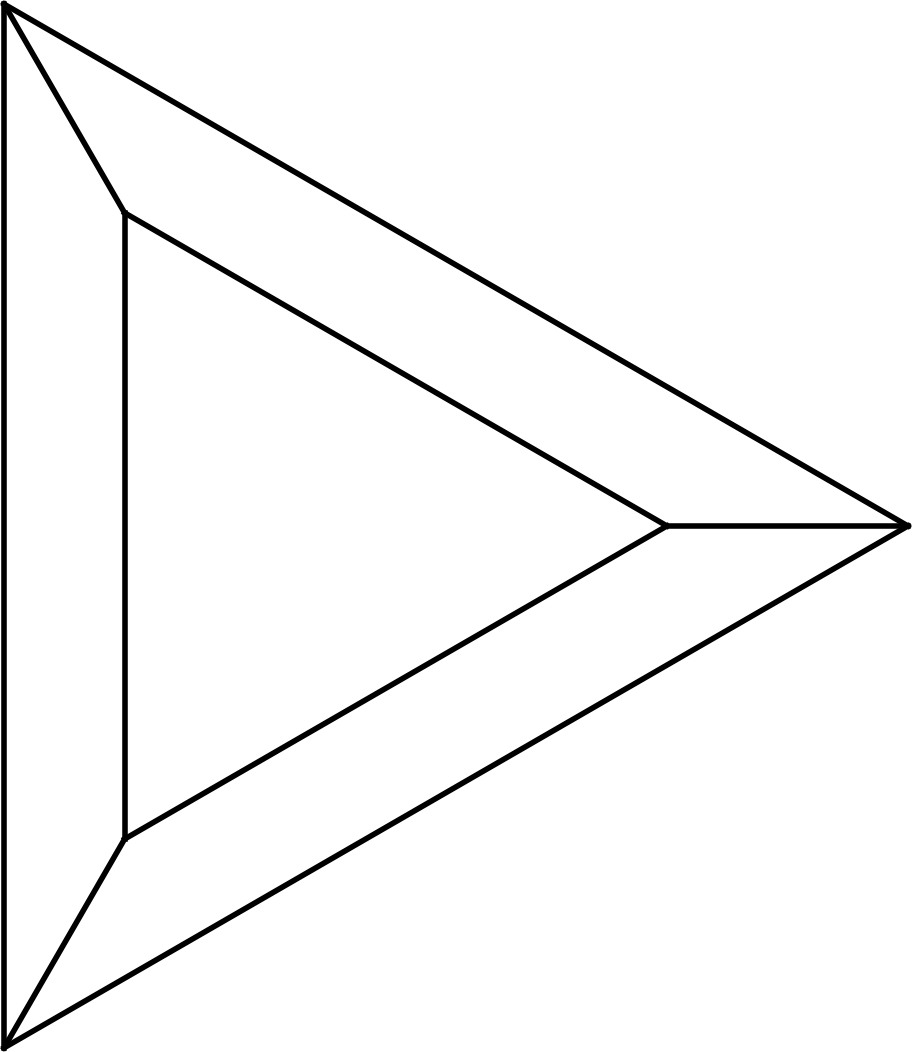}
\caption{Exterior face divided into three isosceles trapezoids.}
\end{figure}
Then we have
\begin{equation}
V=20\qquad,\qquad E=48\qquad,\qquad F=24.
\end{equation}
So that $\chi=-4$ and $g=3$.\\
The initial configuration of the cluster is defined by four parameters. We consider an outer tetrahedron $T$ whose faces are the outer interfaces of the outer bubbles. Three other tetrahedra $T_{\mathrm{i}},T_{\mathrm{m}},T_{\mathrm{e}}$ are defined by scaling the latter with factors $s_{\mathrm{i}}<s_{\mathrm{m}}<s_{\mathrm{e}}$. The tetrahedra $T_{\mathrm{e}},T$ define the outer bubbles. The tetrahedron $T_{\mathrm{i}}$ defines the center bubble. The tetrahedra $T_{\mathrm{i}},T_{\mathrm{e}}$ define the triple torus bubble. The inner double bubbles are defined by right triangular prisms whose bases are the faces of $T_{\mathrm{i}}$ scaled with a factor $s_{\mathrm{f}}$. The tetrahedron $T_{\mathrm{m}}$ is used to define the interfaces shared by the inner double bubbles as the intersection with the right triangle prisms. The different parameters with are graphically described in the following figure.
\begin{figure}[H]
\centering
\includegraphics[scale=0.4]{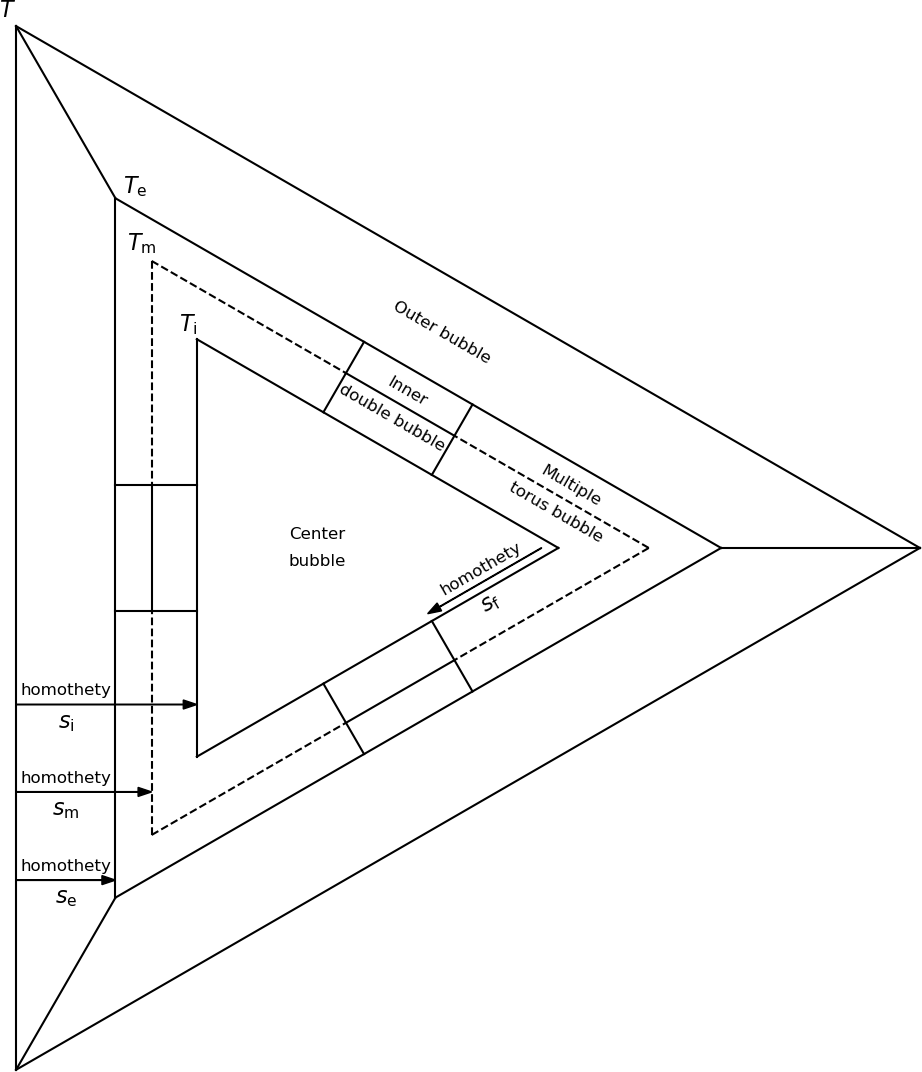}
\caption{$2$D scheme of the configuration.}
\end{figure}
The vertices of the outer regular tetrahedron have coordinates
\begin{equation}
\frac{1}{2}\begin{pmatrix}-1\\-1\\-1\end{pmatrix}\quad,\quad\frac{1}{2}\begin{pmatrix}-1\\1\\1\end{pmatrix}\quad,\quad\frac{1}{2}\begin{pmatrix}1\\-1\\1\end{pmatrix}\quad,\quad\frac{1}{2}\begin{pmatrix}1\\1\\-1\end{pmatrix}.
\end{equation}
When the inner double bubbles are too large, the cluster does not seem to be stable. Hence we choose the following parameters for the initial configuration.
\begin{equation}
s_{\mathrm{i}}=0.25\quad,\quad s_{\mathrm{m}}=0.33\quad,\quad s_{\mathrm{e}}=0.47\quad,\quad s_{\mathrm{f}}=0.6.
\end{equation}
\subsubsection{Energy minimizing cluster}
After minimization  we get the following cluster configuration (total area $3.78$). As for the starting one, we start with showing the outer bubbles, then hide one of them, afterwards we also hide the triple torus bubble and finally we also hide the inner double bubble below the hidden outer bubble.
\begin{figure}[H]
\centering
\includegraphics[scale=0.11]{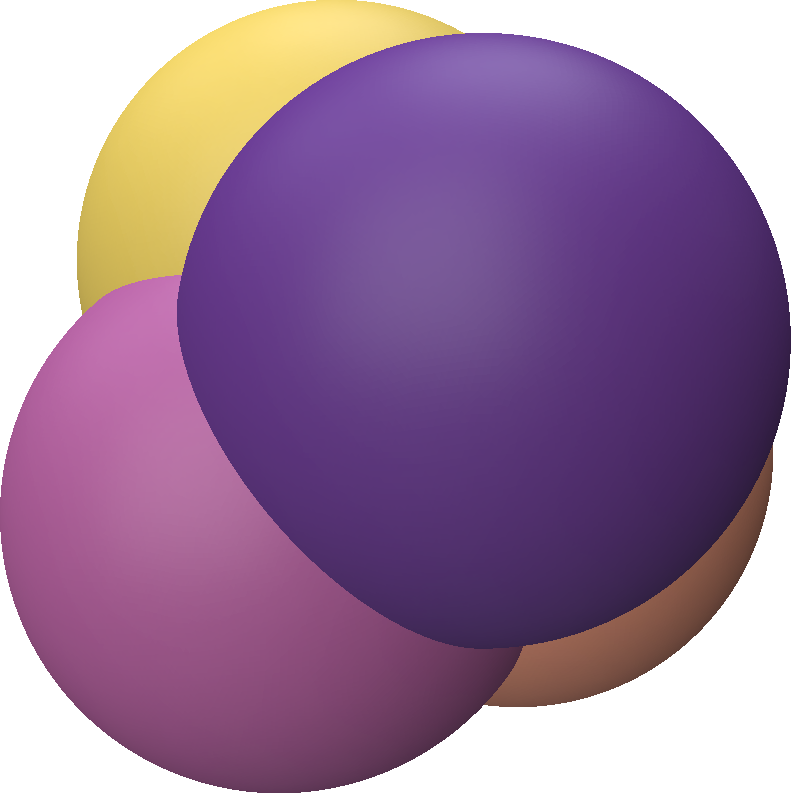}\includegraphics[scale=0.11]{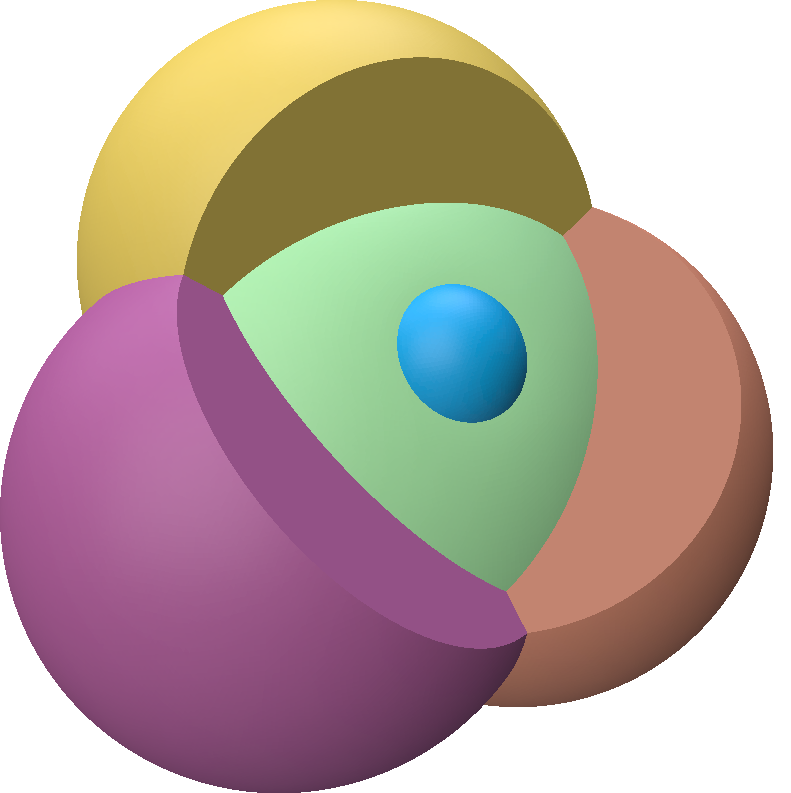}\includegraphics[scale=0.11]{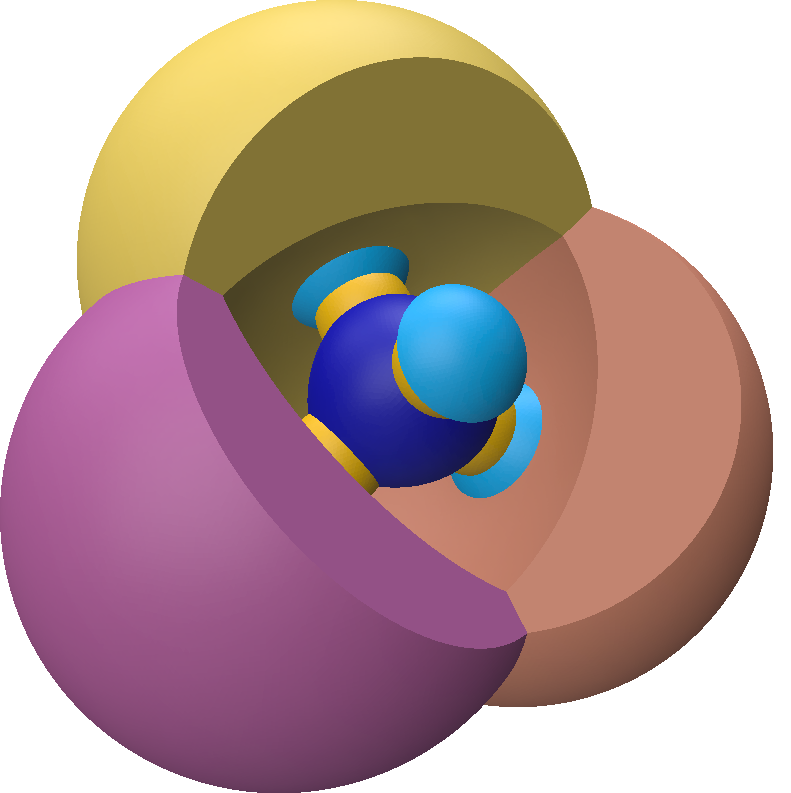}\includegraphics[scale=0.11]{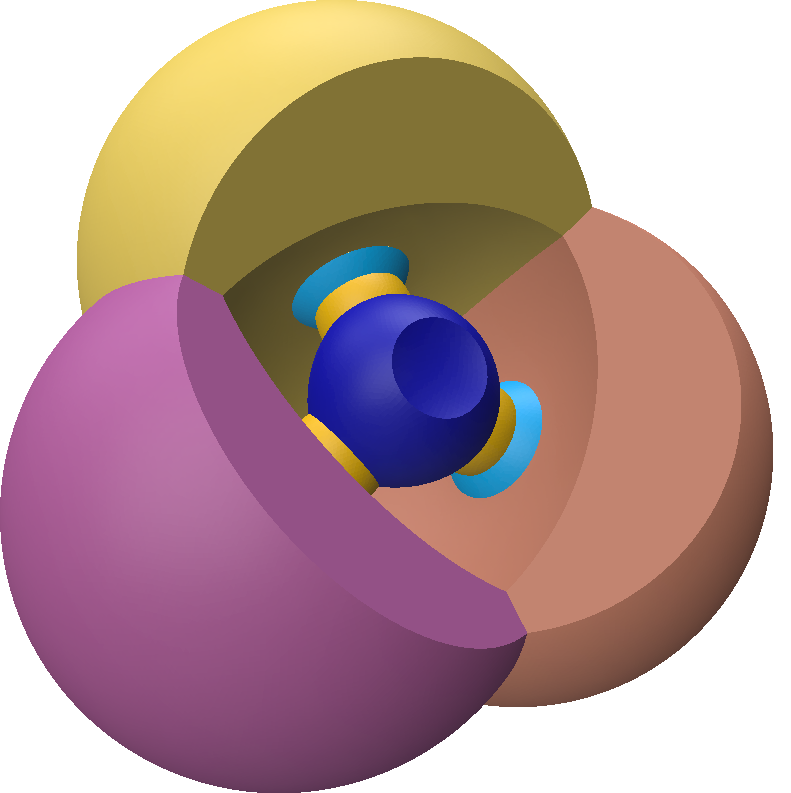}
\caption{Different views of the final cluster.}
\end{figure}
\begin{figure}[H]
\centering
\includegraphics[scale=0.23]{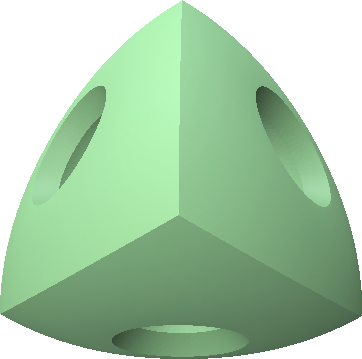}\hspace{2cm}\includegraphics[scale=0.23]{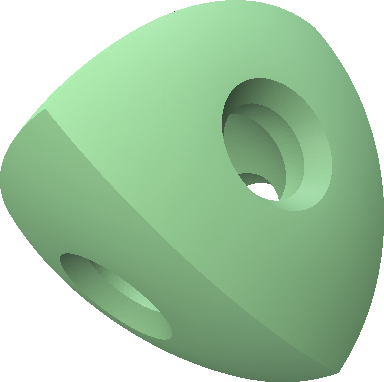}
\caption{Different views of the triple torus bubble.}
\end{figure}
The tetrahedron symmetry seems to be conserved by the minimizing cluster.\\
The inner double bubbles interfaces shared with the triple torus bubble have geometries requiring fine enough discretizations, hence we consider the following two coarsest ones.
\begin{figure}[H]
\centering
\includegraphics[scale=0.18]{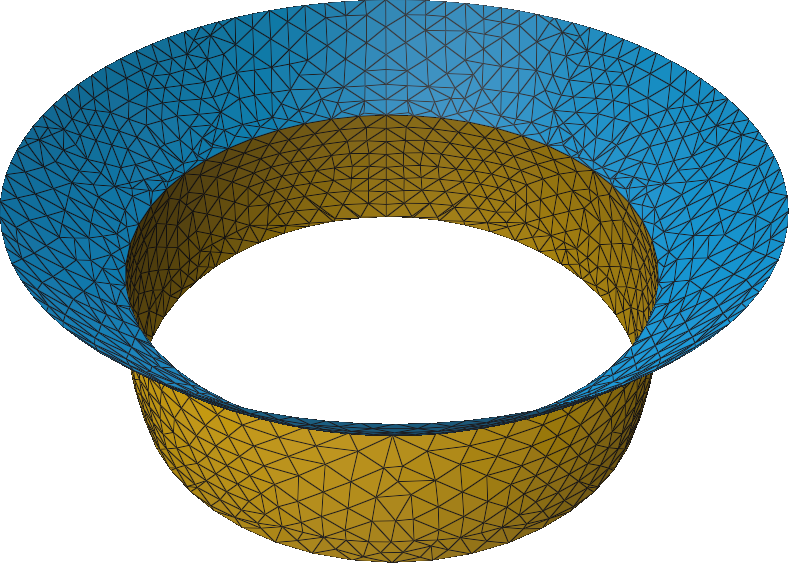}\hspace{2cm}\includegraphics[scale=0.18]{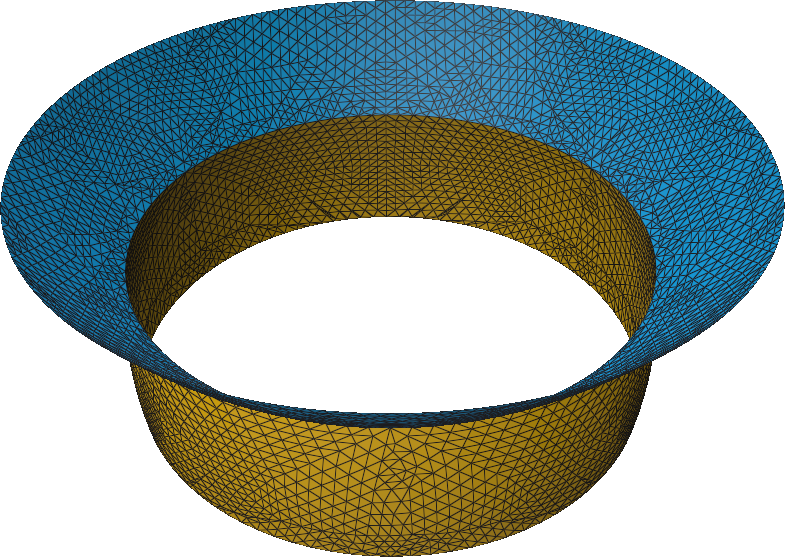}
\caption{Left discretization: $8\times10^{-3}$. Right discretization: $4\times10^{-3}$.}
\end{figure}
The finer discretizations we consider are: $2\times10^{-3}$ and $1.5\times10^{-3}$. For each discretization, the Hessian matrix lowest and highest eigenvalues at the minimum are given in the following table.
\begin{table}[H]
\centering
\begin{tabular}{|c|c|c|}
\hline
Discretization & Lowest eigenvalue & Highest eigenvalue\\
\hline
$8\times10^{-3}$ & $5.0\times10^{-6}$ & $67$\\
\hline
$4\times10^{-3}$ & $6.8\times10^{-7}$ & $81$\\
\hline
$2\times10^{-3}$ & $8.8\times10^{-8}$ & $85$\\
\hline
$1.5\times10^{-3}$ & $4.0\times10^{-8}$ & $95$\\
\hline
\end{tabular}
\caption{Hessian matrix eigenvalues for the different discretizations.}
\end{table}
The previous $14$-bubble cluster with a triple torus bubble seems to be stable. One can thus wonder if more $5$-bubble clusters can be assembled in the same way as \ref{assemble} to build bubbles of higher genus. An obvious idea is to assemble $5$-bubble clusters as \ref{initgen1} around other Platonic solids than the tetrahedron. However, in order to fulfill Plateau's laws, one can only consider the other Platonic solids whose vertices also have valence $3$, namely the cube and the dodecahedron.
\subsection{Fivefold torus bubble}
\subsubsection{Surface Evolver configuration}
Assembling six $5$-bubble cluster \ref{initgen1} around a cube, in a similar way as in section \ref{tetconf}, we get a $20$-bubble cluster with the following configuration.
\begin{figure}[H]
\centering
\includegraphics[scale=0.09]{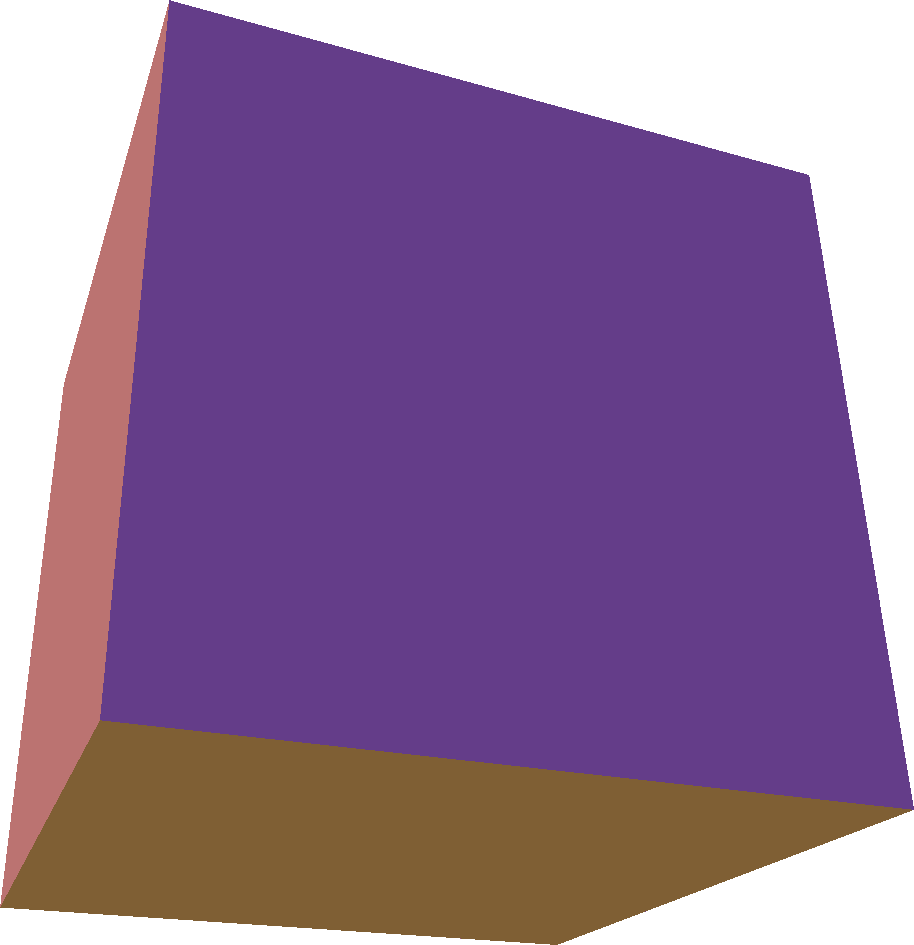}\hspace{0.1cm}\includegraphics[scale=0.09]{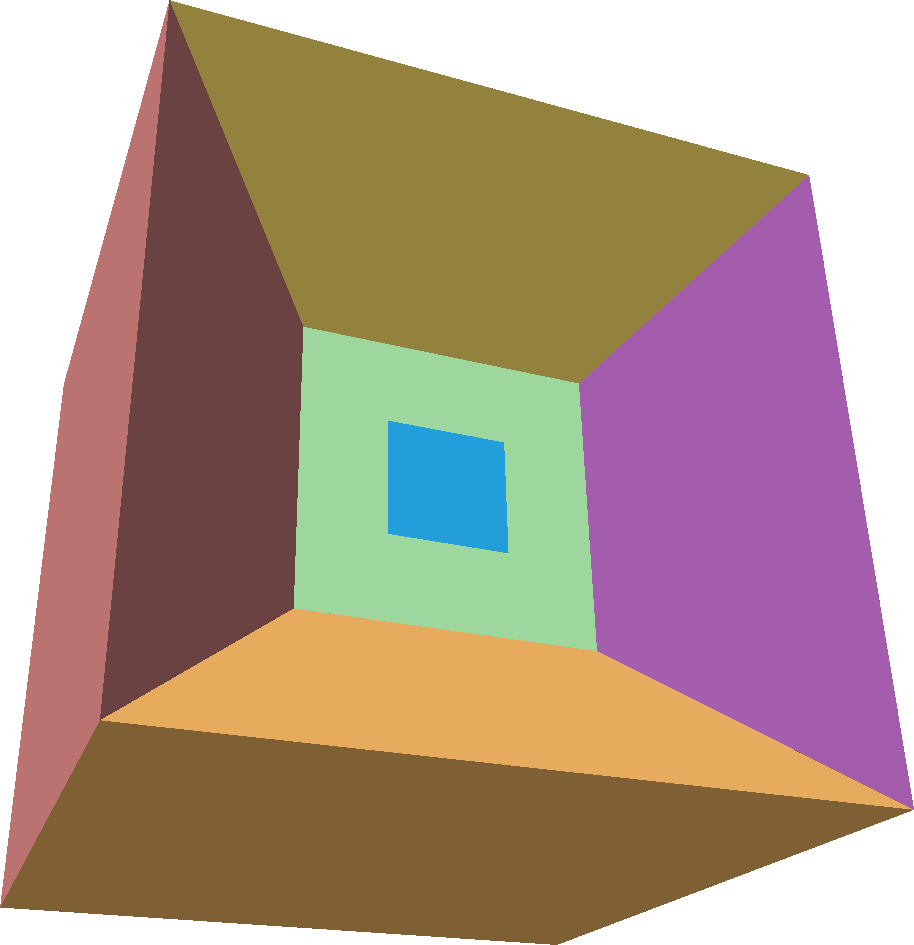}\hspace{0.1cm}\includegraphics[scale=0.09]{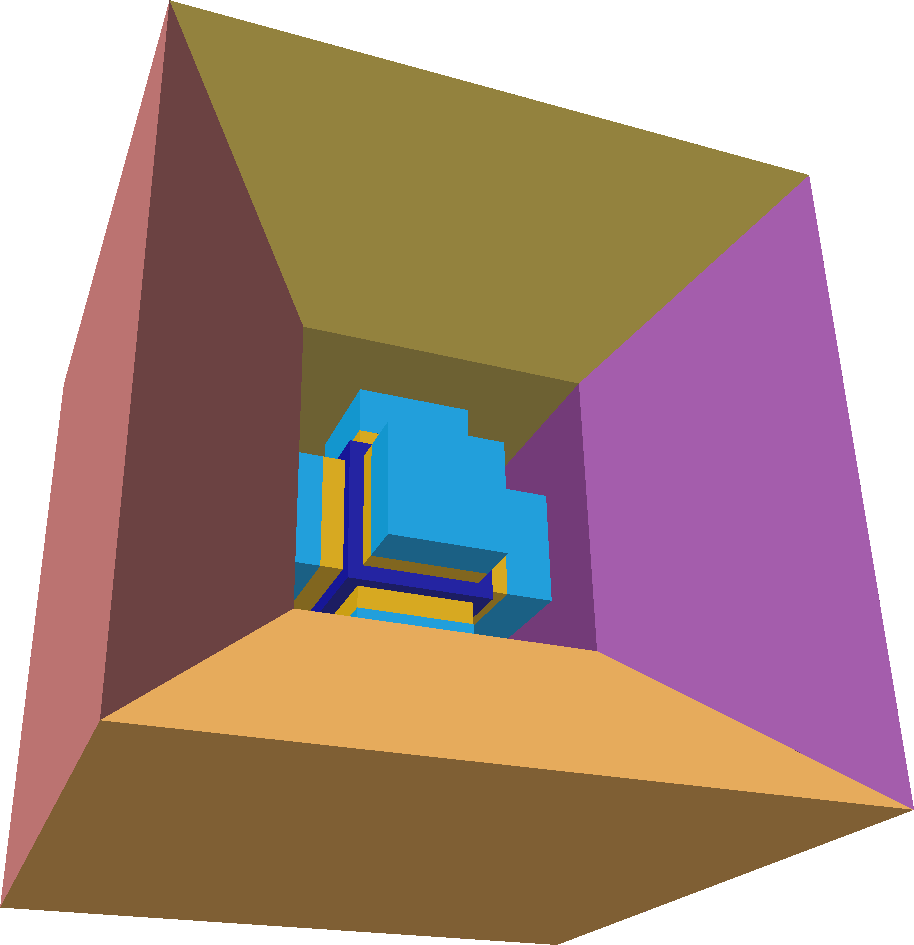}\hspace{0.1cm}\includegraphics[scale=0.09]{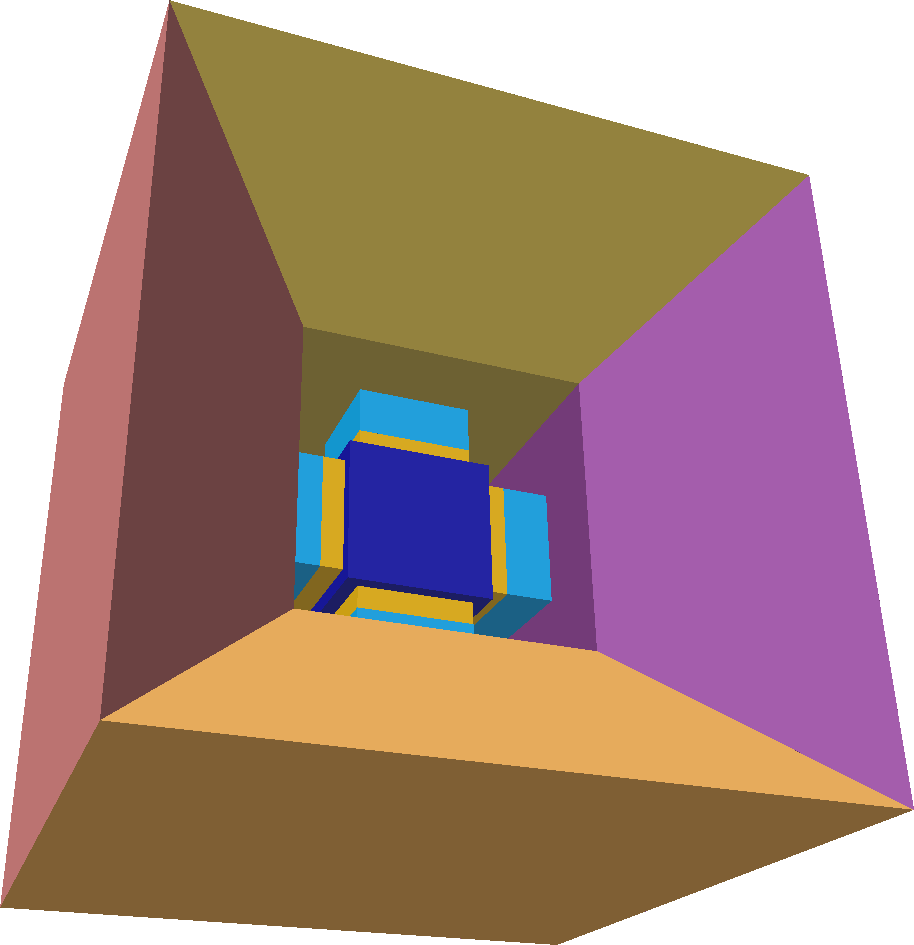}
\caption{Different views of the initial cluster.}
\end{figure}
\begin{figure}[H]
\centering
\includegraphics[scale=0.25]{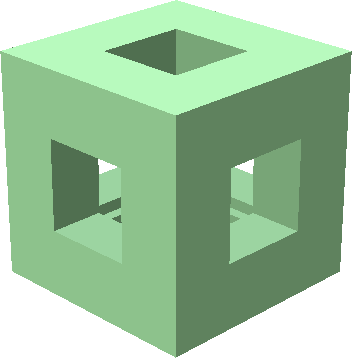}
\caption{Fivefold torus bubble.}
\label{cumultor}
\end{figure}
In the same way as the excavated tetrahedron bubble of figure \ref{tetmultor}, the bubble of figure \ref{cumultor} has genus $5$, i.e. is topologically a fivefold torus.\\
The cluster is parameterized as in section \ref{tetconf} with the outer cube $[-1/2,1/2]^3$. To avoid possible instabilities, the scaling factors have been chosen to avoid large inner double bubbles. They have the following values.
\begin{equation}
s_{\mathrm{i}}=0.23\quad,\quad s_{\mathrm{m}}=0.3\quad,\quad s_{\mathrm{e}}=0.45\quad,\quad s_{\mathrm{f}}=0.8.
\end{equation}
\subsubsection{Energy minimizing cluster}
After minimization, we get the following cluster configuration (total area $8.47$).
\begin{figure}[H]
\centering
\includegraphics[scale=0.09]{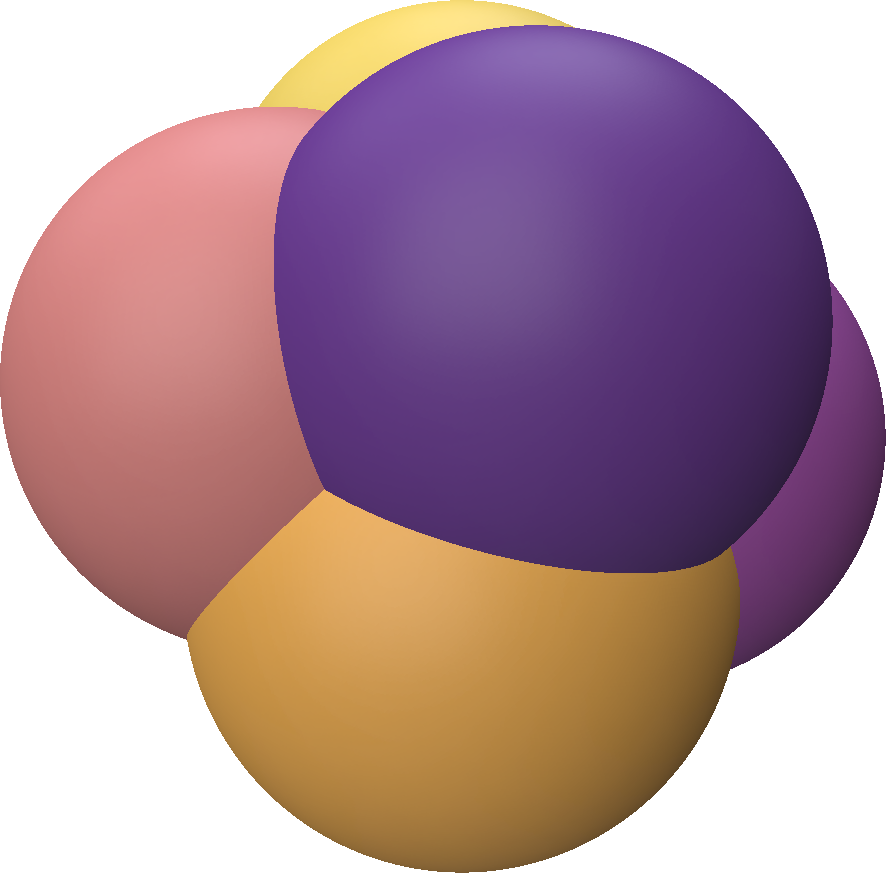}\hspace{0.0cm}\includegraphics[scale=0.09]{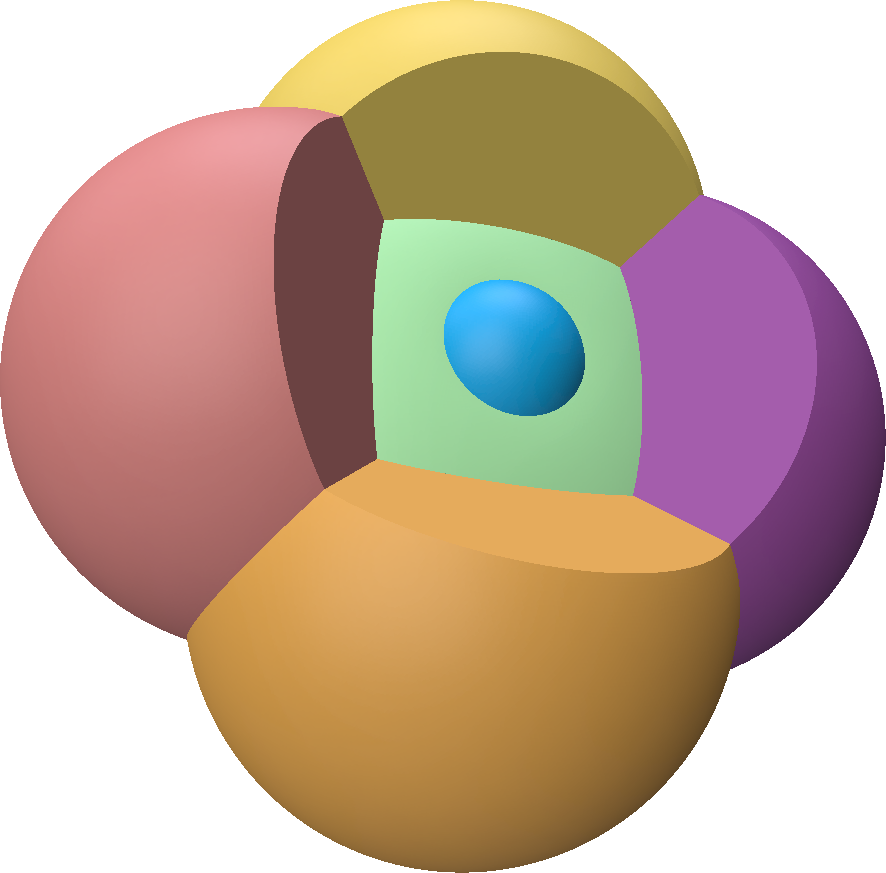}\hspace{0.cm}\includegraphics[scale=0.09]{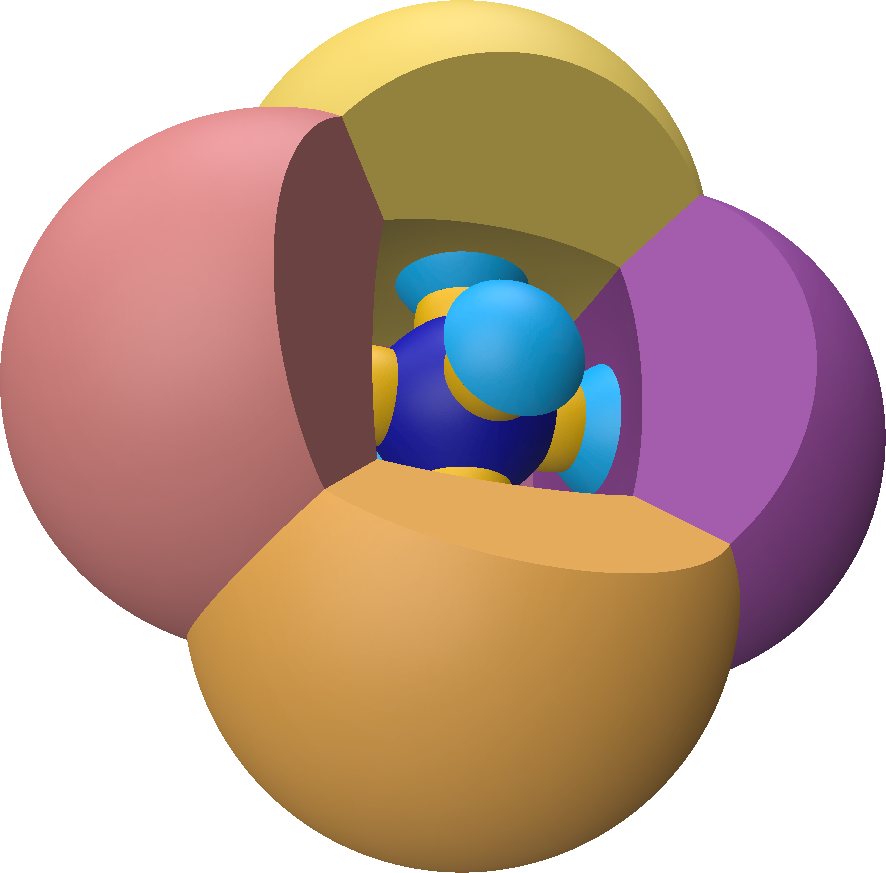}\hspace{0.cm}\includegraphics[scale=0.09]{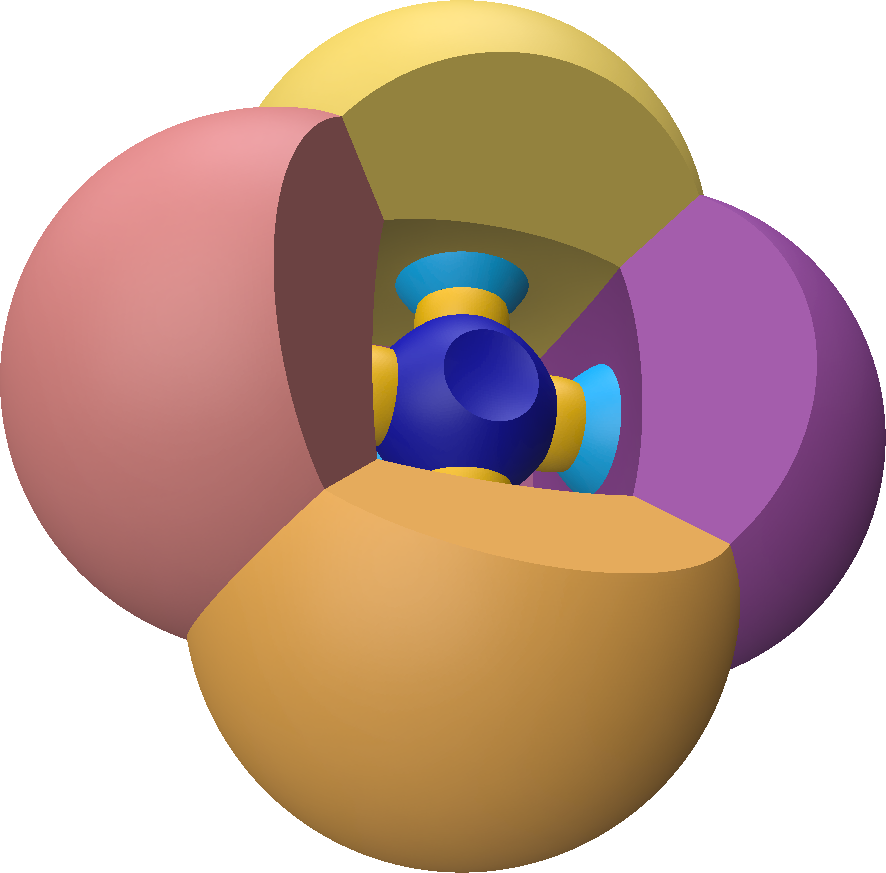}
\caption{Different views of the final cluster.}
\end{figure}
\begin{figure}[H]
\centering
\includegraphics[scale=0.2]{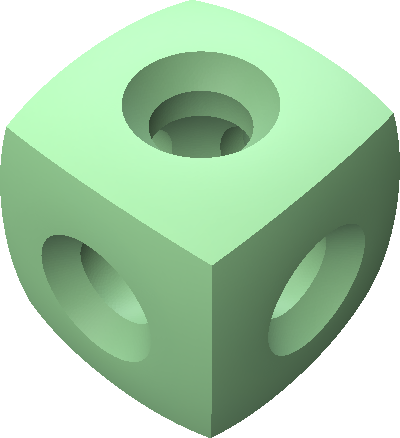}\hspace{1cm}\includegraphics[scale=0.2]{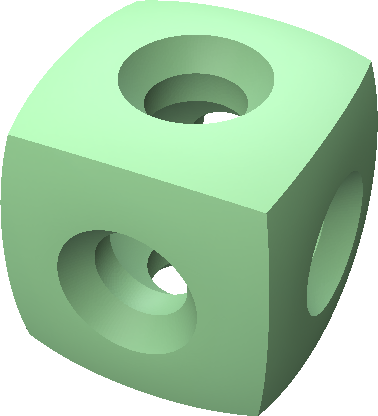}\hspace{1cm}\includegraphics[scale=0.2]{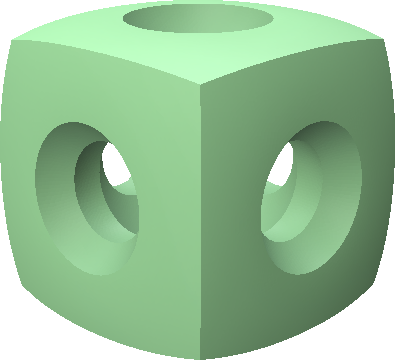}
\caption{Different views of the fivefold torus bubble.}
\end{figure}
Similarly to the previous cluster, the cube symmetry seems to be conserved.\\
The interfaces of the double bubbles shared with the fivefold torus bubble require again a fine discretization. The two coarsest ones we chose are plotted in the following figures.
\begin{figure}[H]
\centering
\includegraphics[scale=0.15]{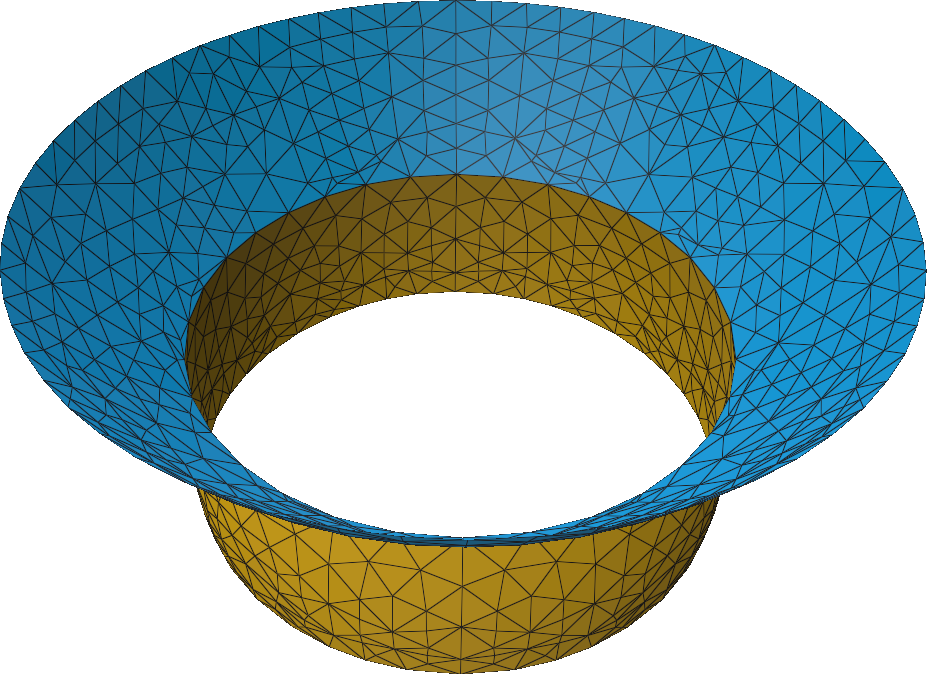}\hspace{2cm}\includegraphics[scale=0.15]{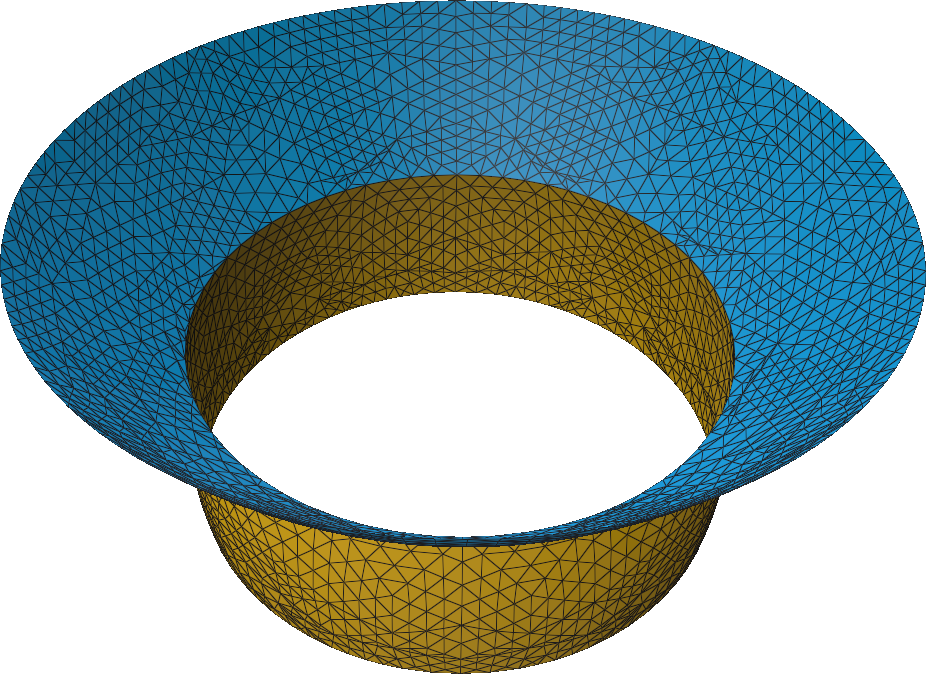}
\caption{Left discretization: $8\times10^{-3}$. Right discretization: $4\times10^{-3}$.}
\end{figure}
Two finer discretizations are also considered: $3\times10^{-3}$ and $2\times10^{-3}$. For each discretization, the Hessian matrix lowest and highest eigenvalues at the minimum are given in the following table.
\begin{table}[H]
\centering
\begin{tabular}{|c|c|c|}
\hline
Discretization & Lowest eigenvalue & Highest eigenvalue\\
\hline
$8\times10^{-3}$ & $2.3\times10^{-6}$ & $25$\\
\hline
$4\times10^{-3}$ & $3.0\times10^{-7}$ & $25$\\
\hline
$3\times10^{-3}$ & $1.3\times10^{-7}$ & $26$\\
\hline
$2\times10^{-3}$ & $3.8\times10^{-8}$ & $27$\\
\hline
\end{tabular}
\caption{Hessian matrix eigenvalues for the different discretizations.}
\end{table}
As the previous clusters, this one seems to be stable as well.
\subsection{Elevenfold torus bubble}
\subsubsection{Surface Evolver configuration}
Assembling now twelve $5$-bubble cluster \ref{initgen1} around a dodecahedron, in a similar way as in section \ref{tetconf}, we get a $38$-bubble cluster with the following starting configuration.
\begin{figure}[H]
\centering
\includegraphics[scale=0.08]{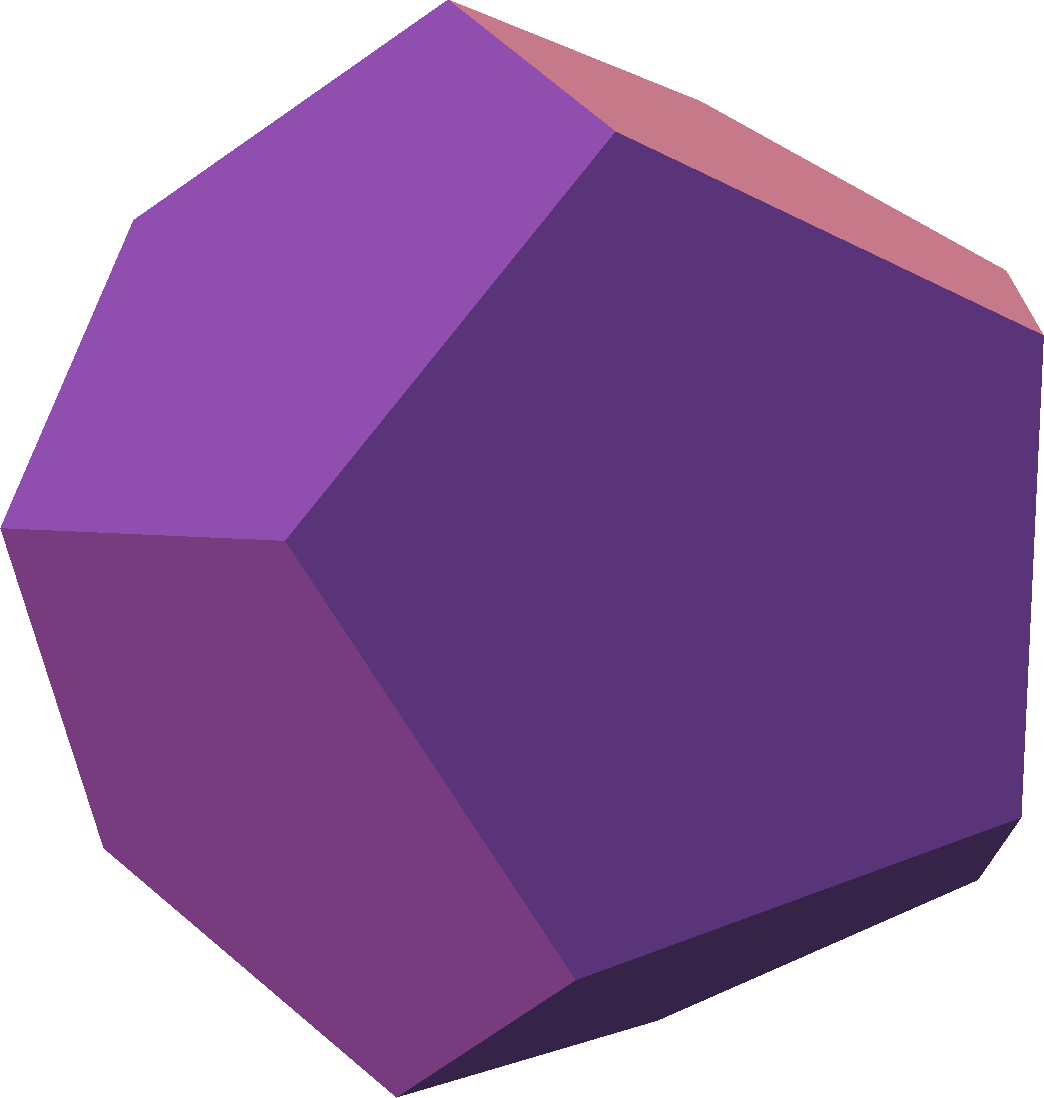}\hspace{0.cm}\includegraphics[scale=0.08]{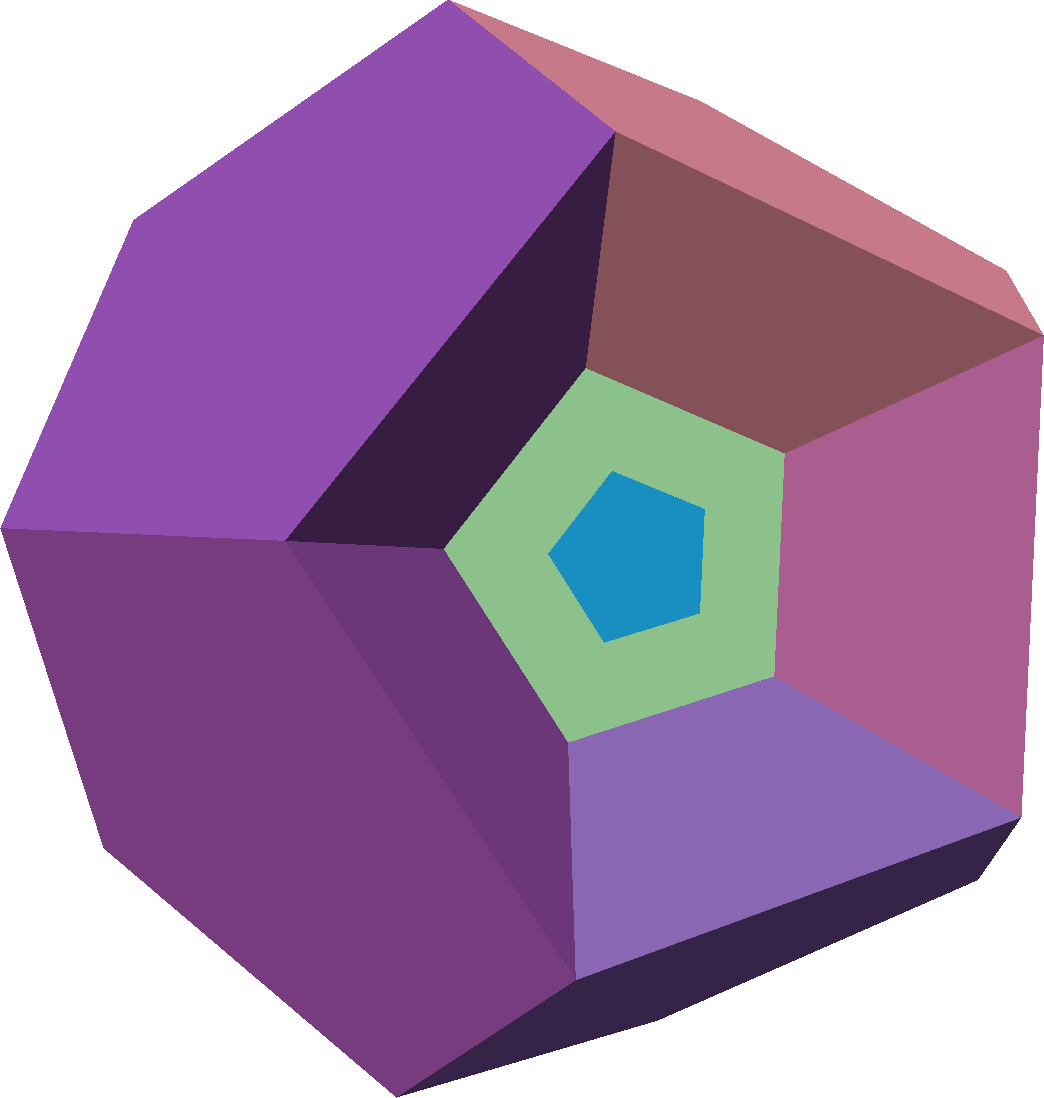}\hspace{0.cm}\includegraphics[scale=0.08]{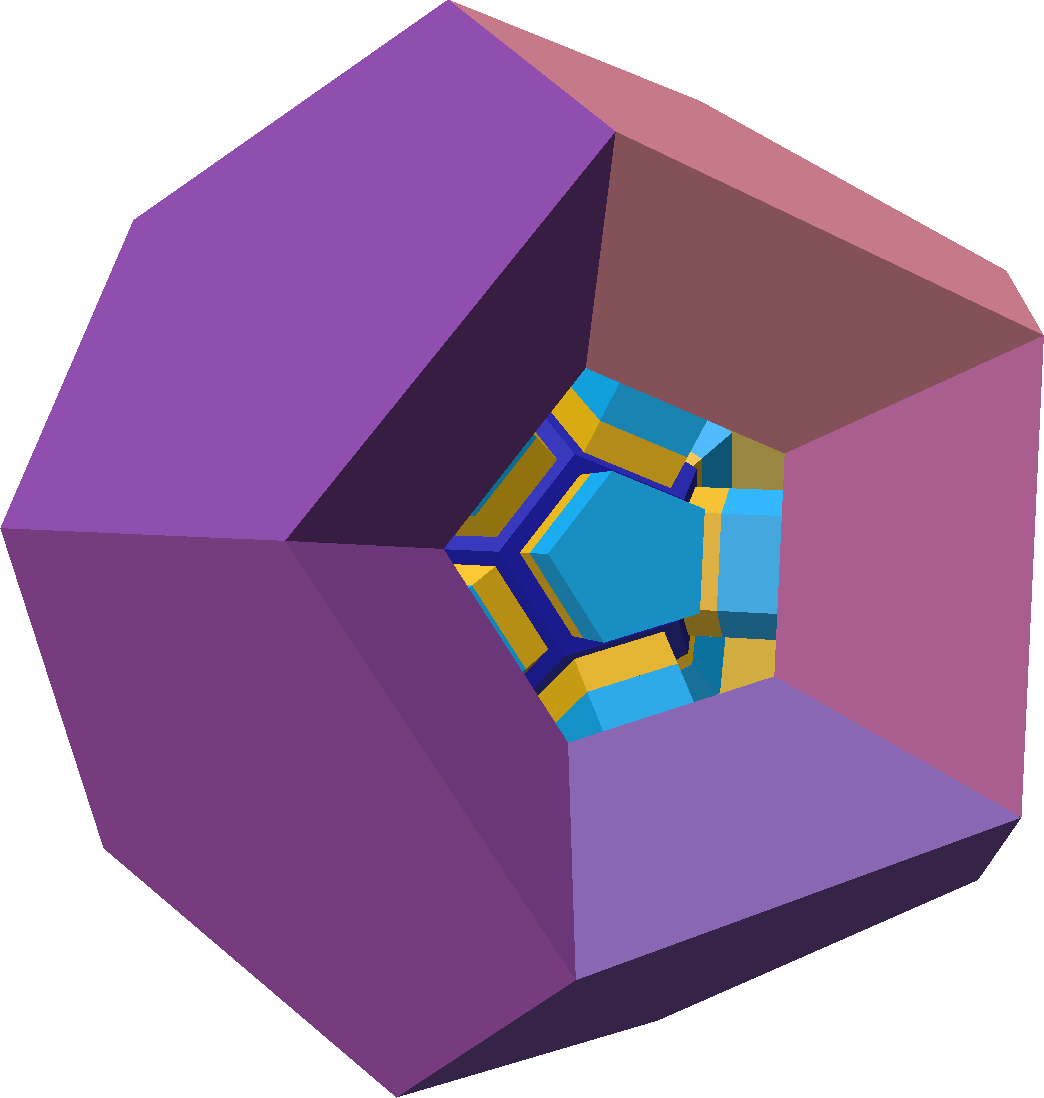}\hspace{0.cm}\includegraphics[scale=0.08]{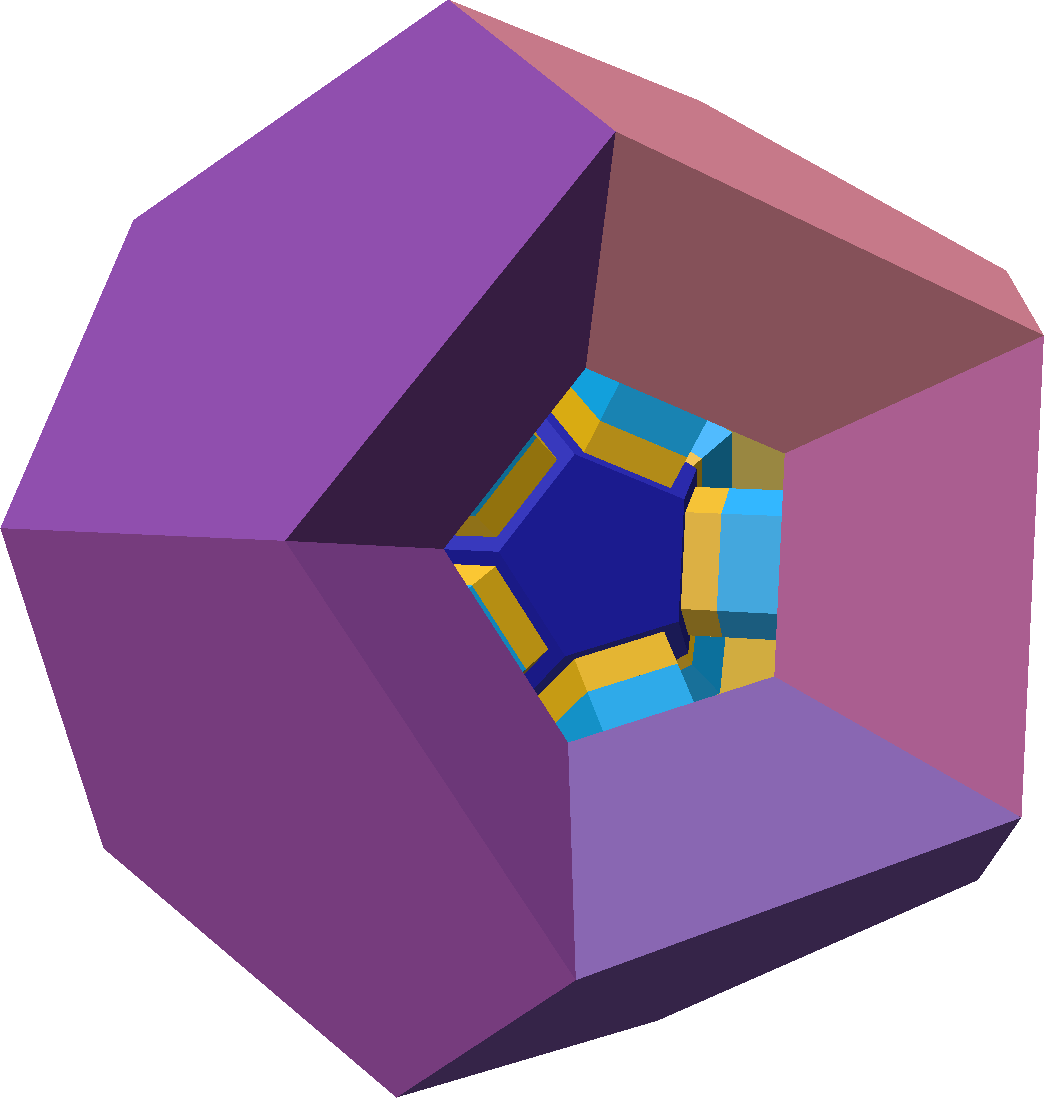}
\caption{Different views of the initial cluster.}
\end{figure}
\begin{figure}[H]
\centering
\includegraphics[scale=0.11]{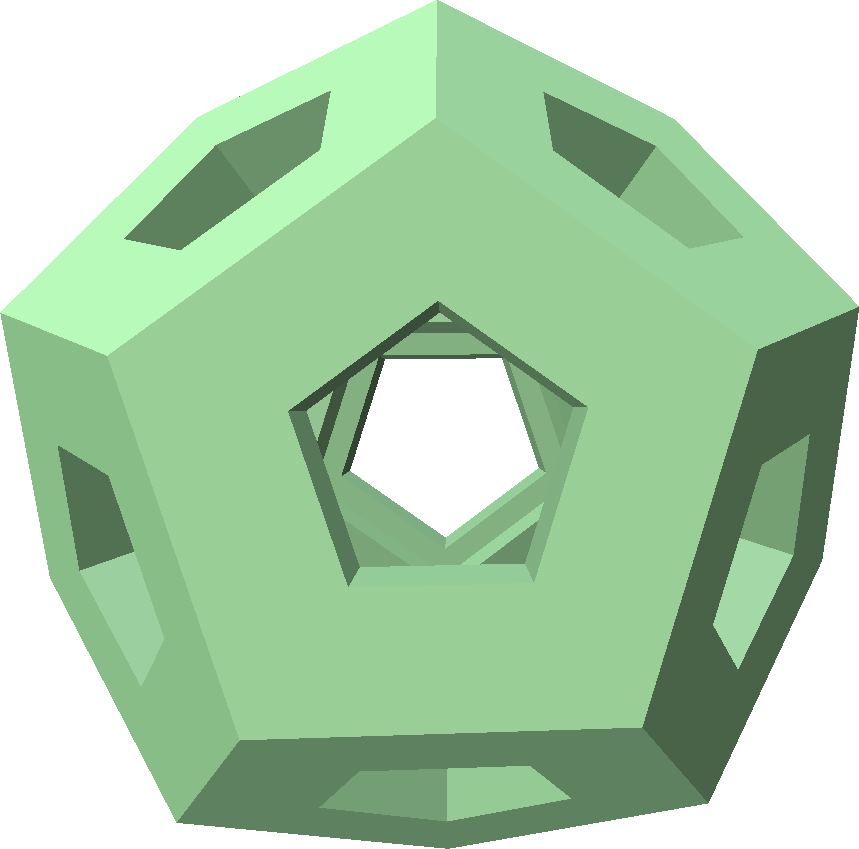}\hspace{2cm}\includegraphics[scale=0.1]{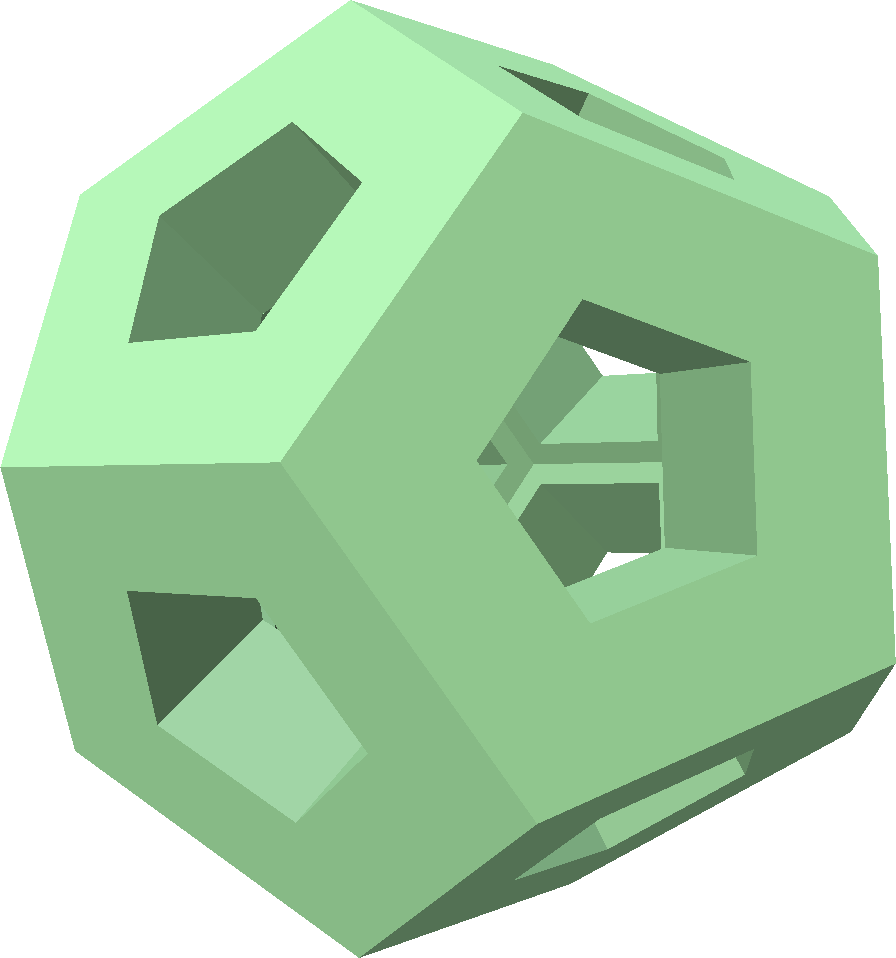}
\caption{Different views of the elevenfold torus bubble.}
\label{dodmultor}
\end{figure}
The bubble of the previous figure \ref{dodmultor} has genus $11$, i.e. is topologically an elevenfold torus.\\
The cluster is parameterized as in the previous sections. The vertices of the outer regular dodecahedron have coordinates
\begin{equation}
\frac{1}{\sqrt{3}}\begin{pmatrix}\pm1\\\pm1\\\pm1\end{pmatrix},
\end{equation}
And all cyclic permutations of
\begin{equation}
\frac{1}{\sqrt{3}}\begin{pmatrix}0\\\pm\varphi\\\pm(\varphi-1)\end{pmatrix}\quad,\quad\varphi=\frac{1+\sqrt{5}}{2}.
\end{equation}
The scaling factors have the following values.
\begin{equation}
s_{\mathrm{i}}=0.3\quad,\quad s_{\mathrm{m}}=0.38\quad,\quad s_{\mathrm{e}}=0.52\quad,\quad s_{\mathrm{f}}=0.8.
\end{equation}
\subsubsection{Energy minimizing cluster}
After minimization, we get the following cluster configuration (total area $20.23$).
\begin{figure}[H]
\centering
\includegraphics[scale=0.08]{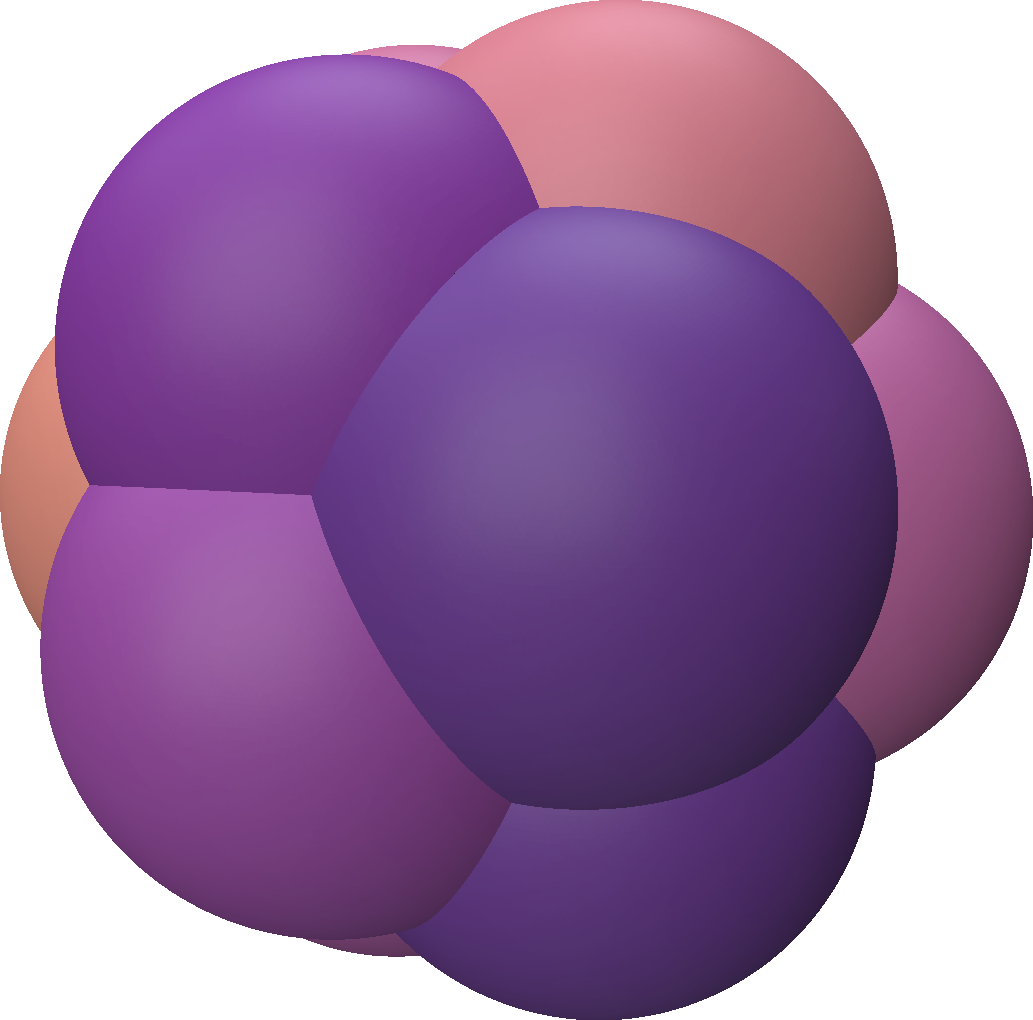}\hspace{0.1cm}\includegraphics[scale=0.08]{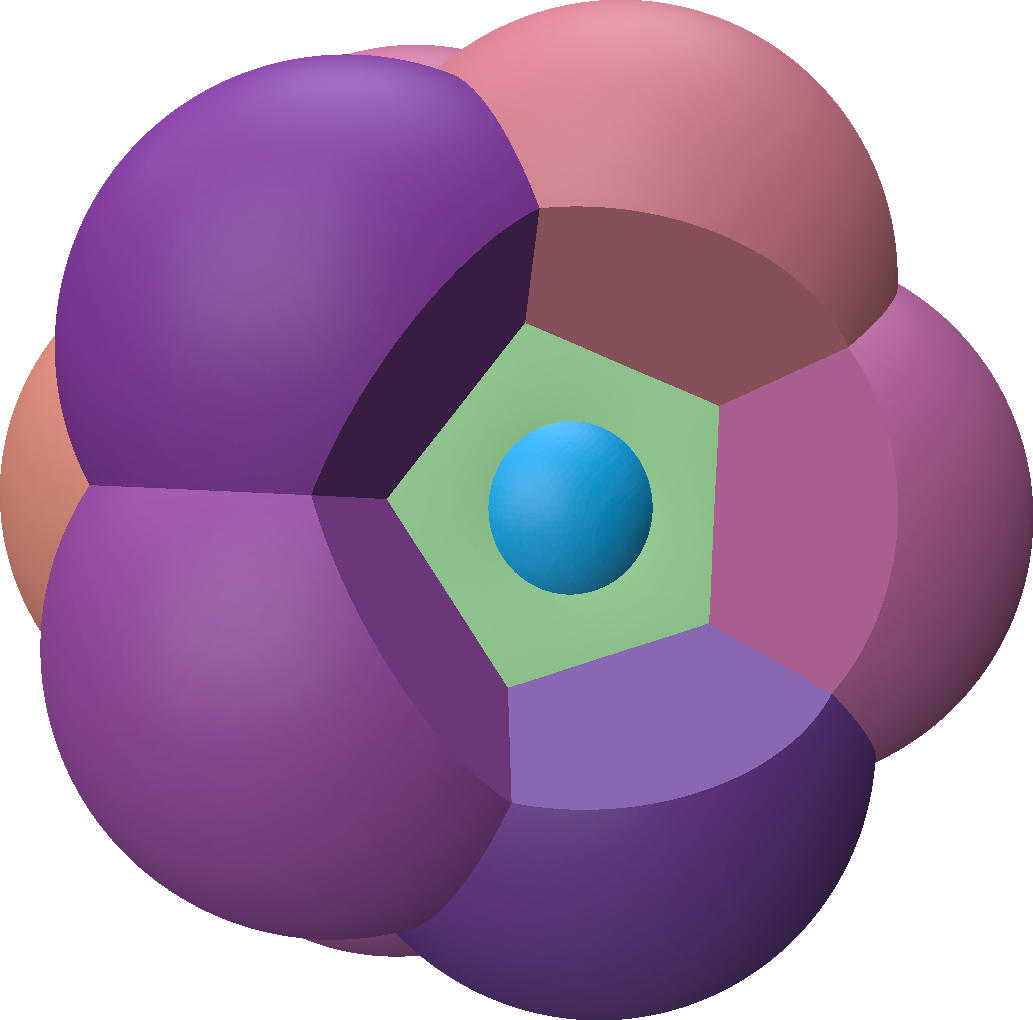}\hspace{0.1cm}\includegraphics[scale=0.08]{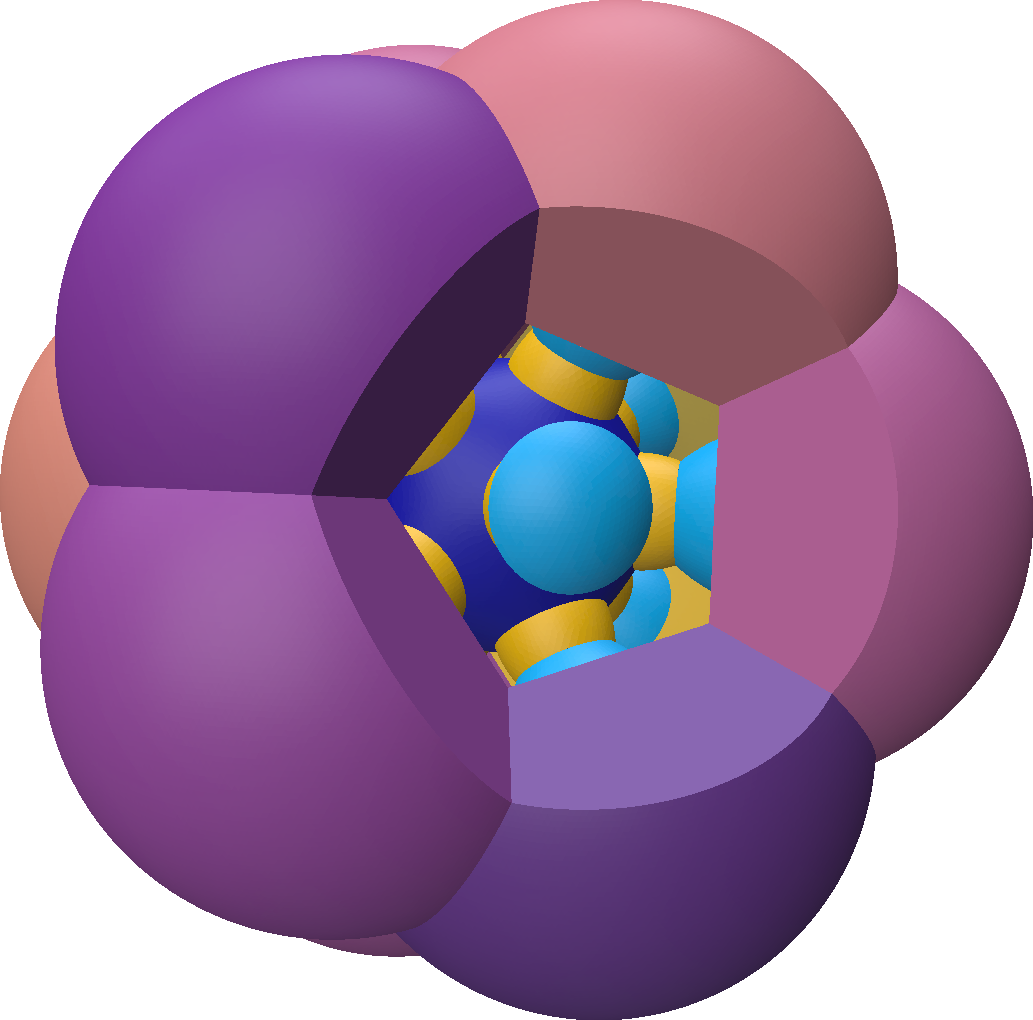}\hspace{0.1cm}\includegraphics[scale=0.08]{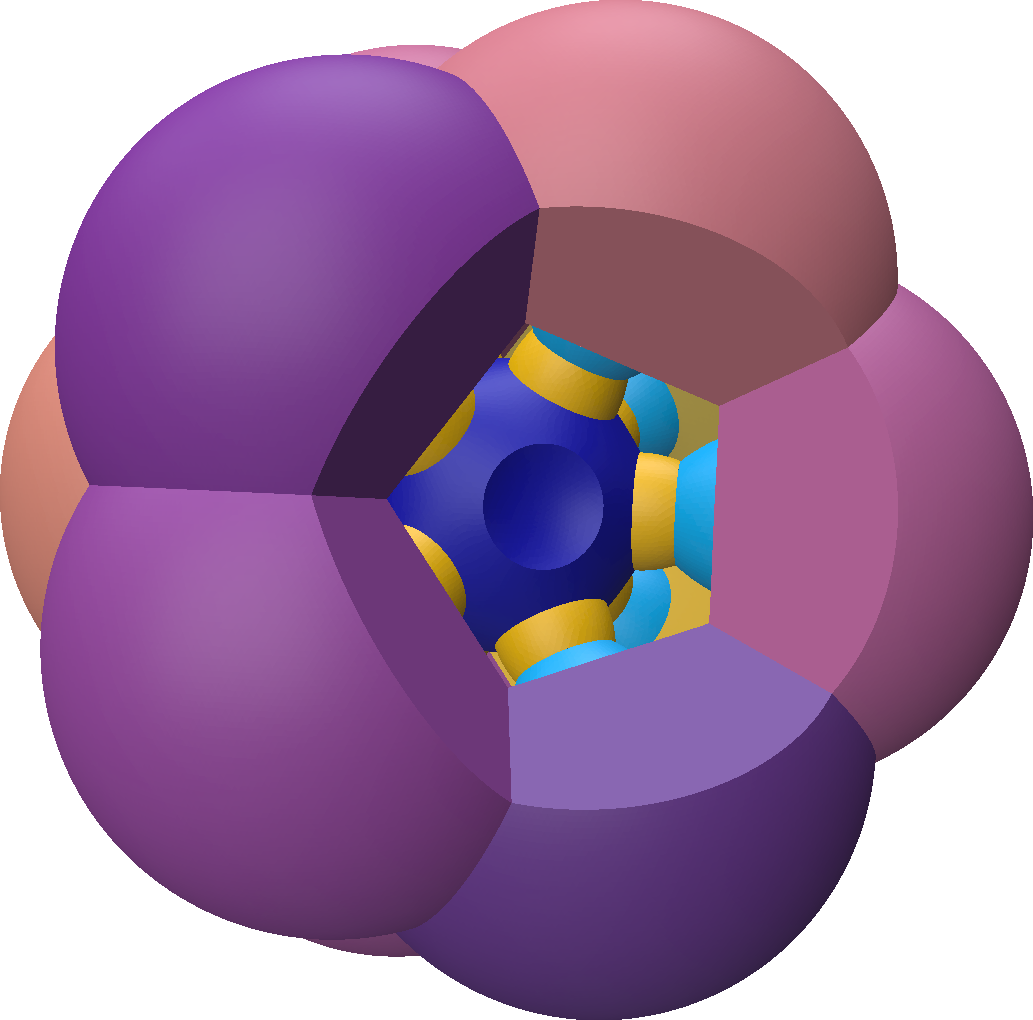}
\caption{Different views of the final cluster.}
\end{figure}
\begin{figure}[H]
\centering
\includegraphics[scale=0.08]{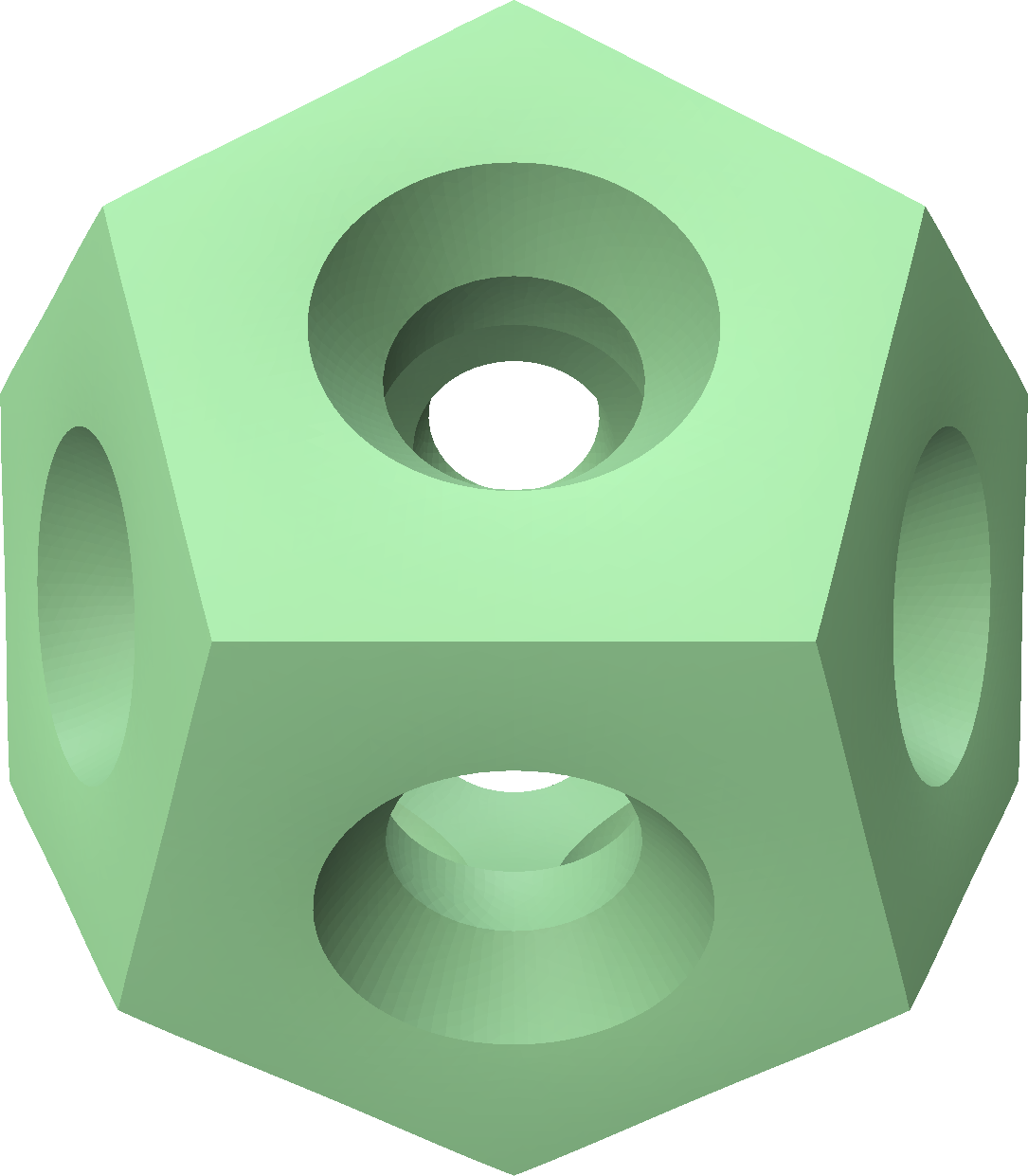}\hspace{2cm}\includegraphics[scale=0.08]{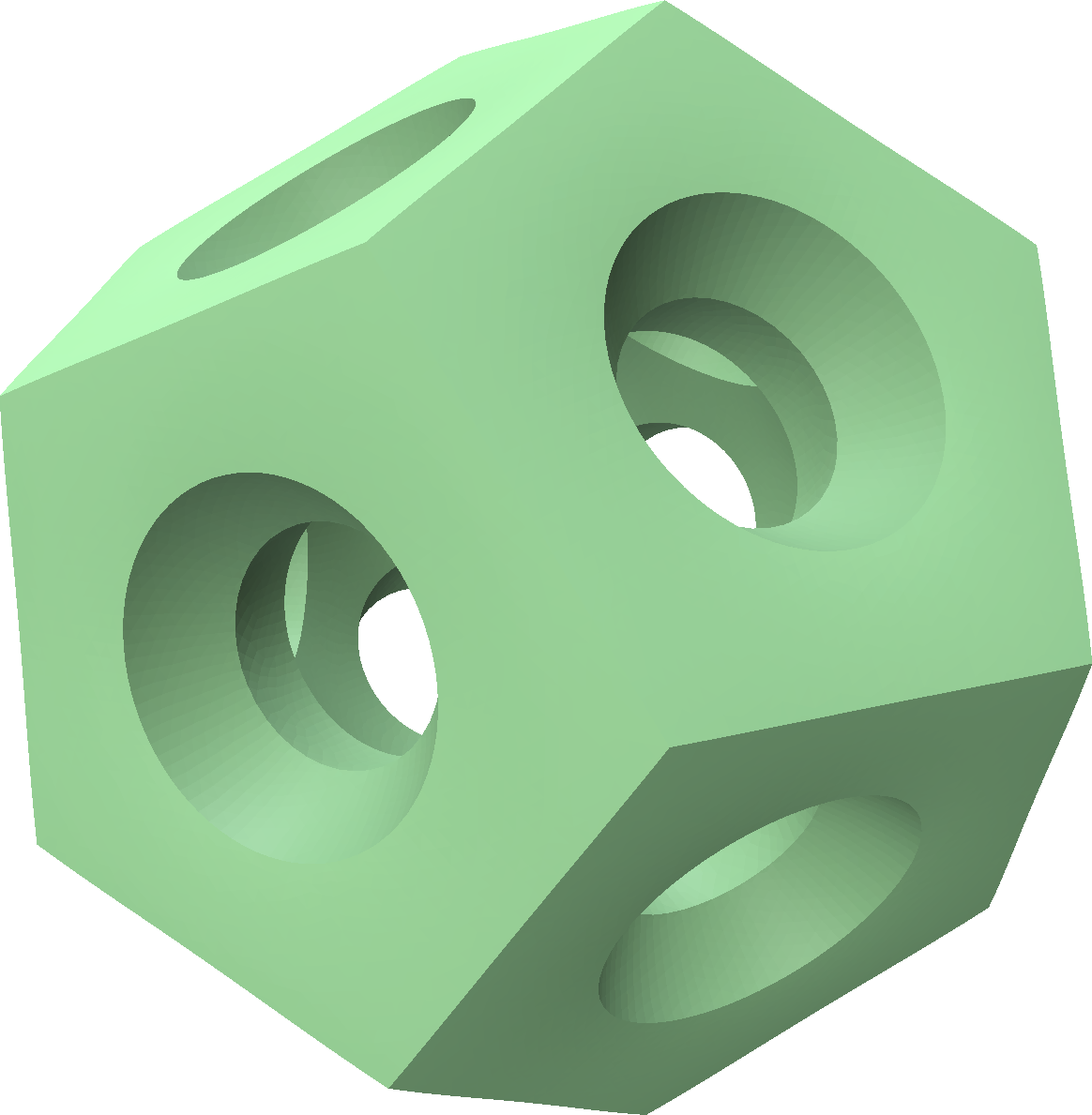}
\caption{Different views of the elevenfold torus bubble.}
\end{figure}
Similarly to the other clusters, the dodecahedron symmetry seems to be conserved.\\
We chose fine enough discretizations to be relevant considering the geometry of the double bubbles interfaces shared with the elevenfold torus bubble. The two coarsest meshes we consider are plotted in the following figures.
\begin{figure}[H]
\centering
\includegraphics[scale=0.12]{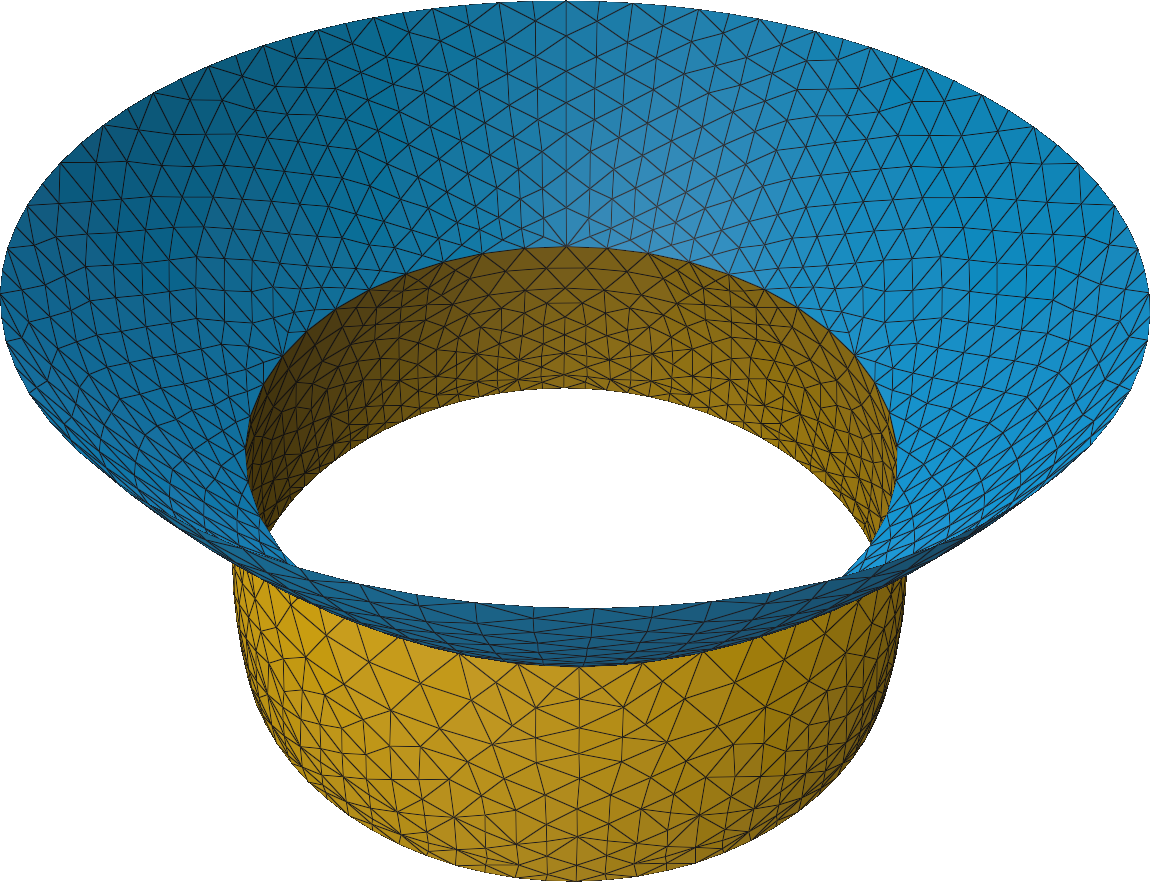}\hspace{2cm}\includegraphics[scale=0.12]{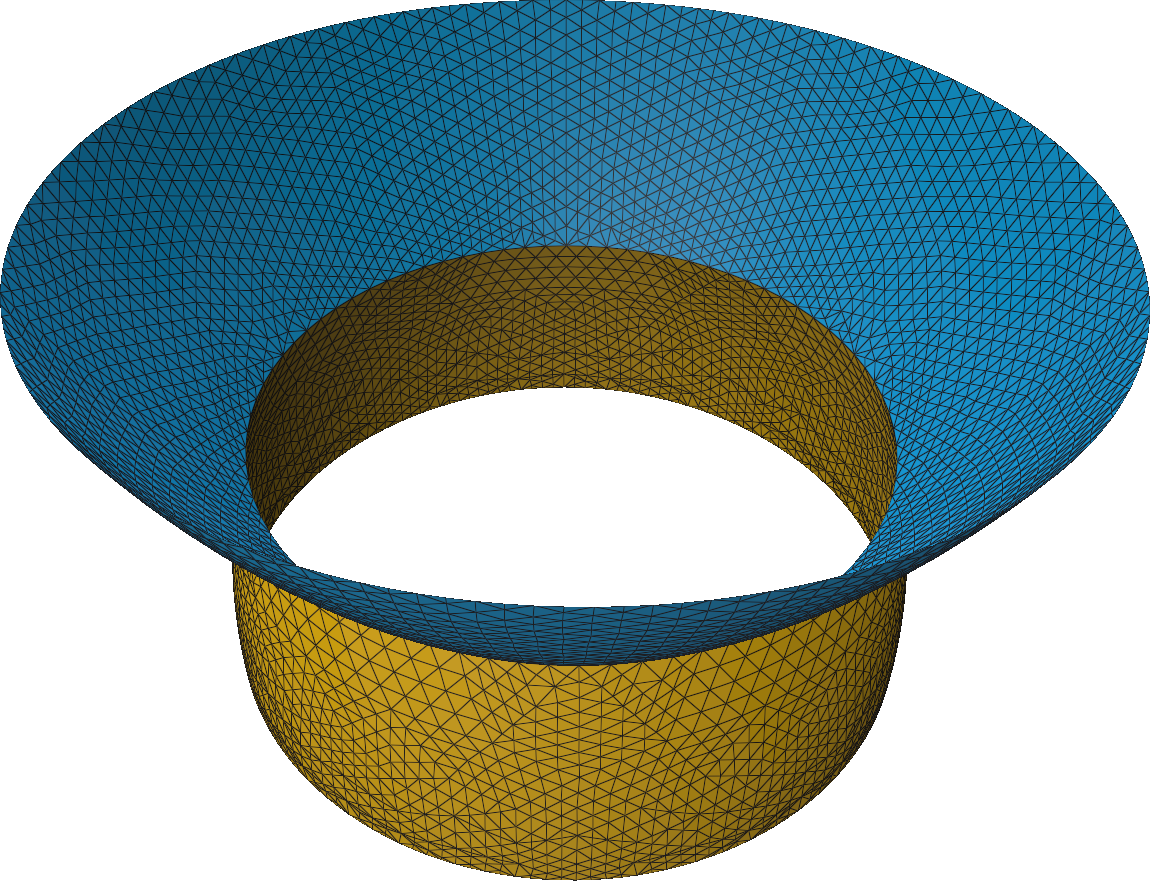}
\caption{Left discretization: $1.6\times10^{-2}$. Right discretization: $8\times10^{-3}$.}
\end{figure}
For this cluster, the center bubble has also some parts which require fine enough discretizations. The previous ones are shown in the following figure.
\begin{figure}[H]
\centering
\includegraphics[scale=0.1]{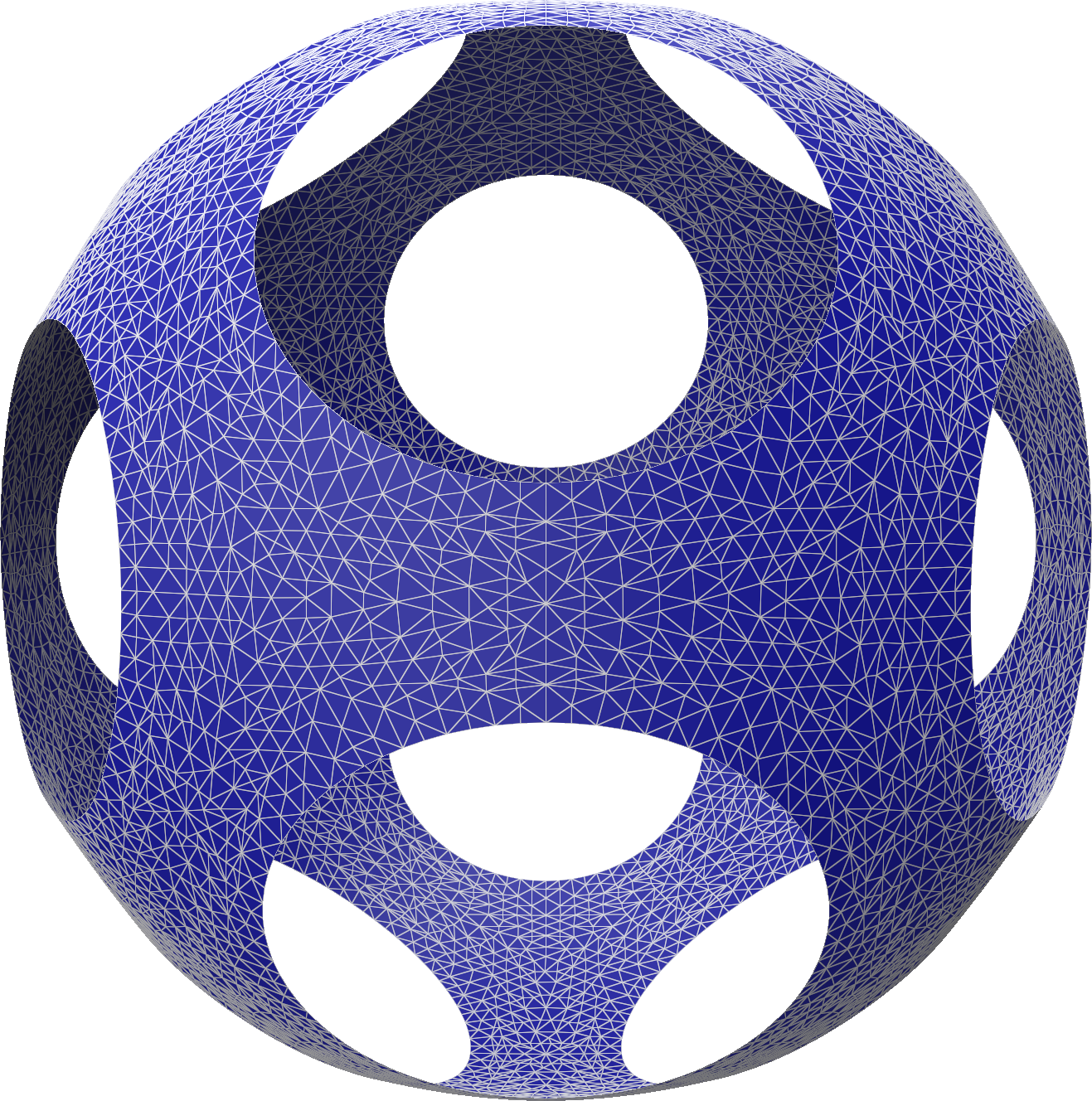}\hspace{2cm}\includegraphics[scale=0.1]{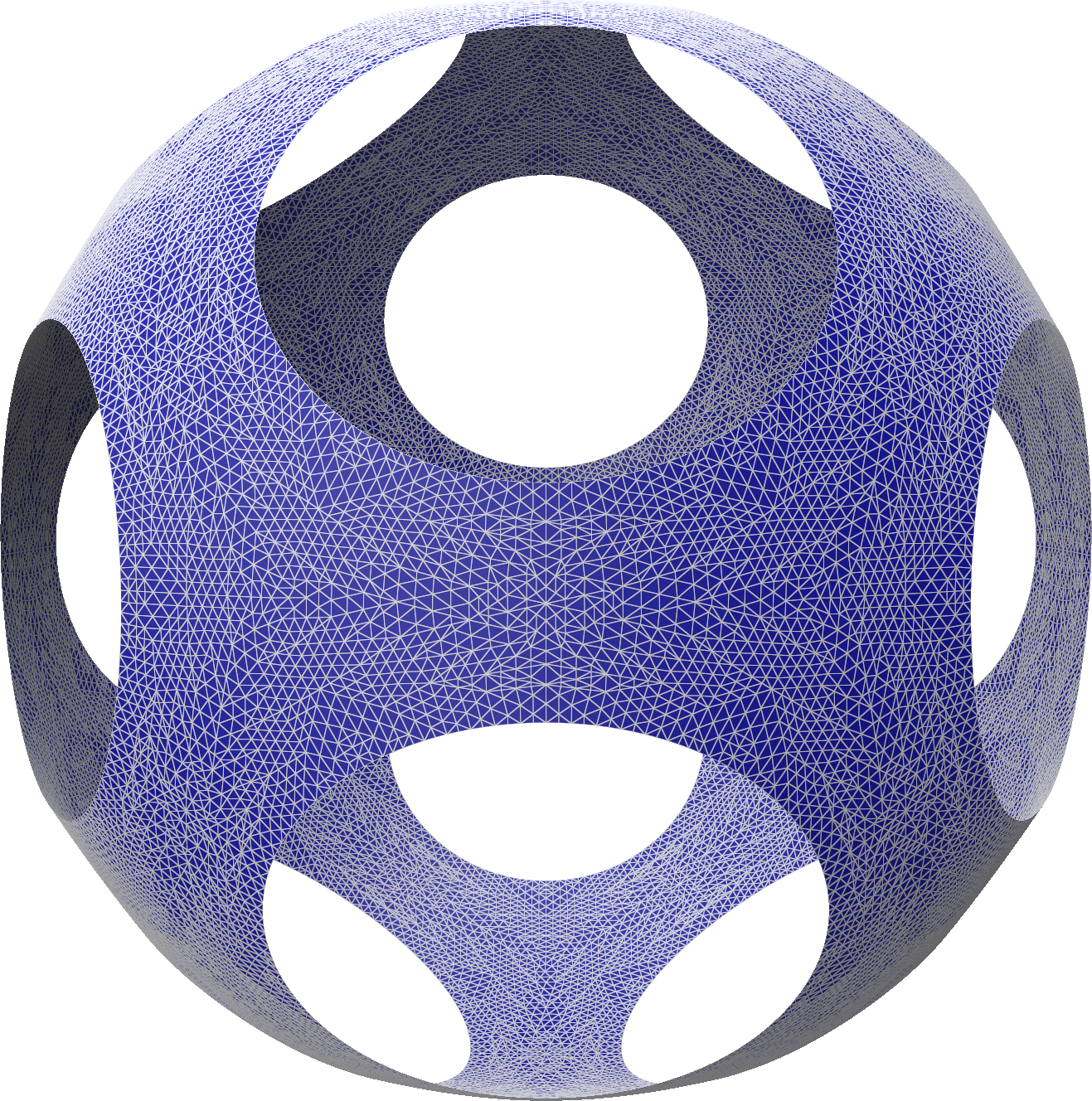}
\caption{Left discretization: $1.6\times10^{-2}$. Right discretization: $8\times10^{-3}$.}
\end{figure}
The finer discretizations we consider are: $4\times10^{-3}$ and $3\times10^{-3}$. For each discretization, the Hessian matrix lowest and highest eigenvalues at the minimum are given in the following table.
\begin{table}[H]
\centering
\begin{tabular}{|c|c|c|}
\hline
Discretization & Lowest eigenvalue & Highest eigenvalue\\
\hline
$1.6\times10^{-2}$ & $9.8\times10^{-6}$ & $25$\\
\hline
$8\times10^{-3}$ & $1.2\times10^{-6}$ & $26$\\
\hline
$4\times10^{-3}$ & $1.5\times10^{-7}$ & $26$\\
\hline
$3\times10^{-3}$ & $1.1\times10^{-7}$ & $26$\\
\hline
\end{tabular}
\caption{Hessian matrix eigenvalues for the different discretizations.}
\end{table}
This cluster seems to be stable as well.
\section{Stability}
Although the command ``longj'' (long jiggle) is ``archaic'' according to Surface Evolver documentation, one can use it for a coarse study of the stability of the clusters. This command moves the vertices of the mesh according to a wavevector, a phase and a vector amplitude. Using random values provided by Surface Evolver, after gradient steps and some mesh polishing, the clusters are back to their previous stable configurations. Though it is not its main purpose, the ``j'' (jiggle) command can be tested as well. In this case, the vertices locations are perturbed according to a Gaussian noise. Using the default ``temperature'' ($0.05$), after some work as previously, the clusters are back to their previous stable configurations.
\section{Data files}
All simulations results as well as Surface Evolver configuration files and Python scripts allowing to create the configurations with some flexibility are available on \href{https://doi.org/10.5281/zenodo.14962654}{Zenodo}. An independent viewer using Polyscope \cite{polyscope} and multimaterial meshes versions (as used by Multitracker \cite{multitracker}) of Surface Evolver dump files are also available.
\section{Conclusion}
With simple numerical examples, we have shown that stable soap bubble clusters can contain multiple torus bubbles. Some improvements of this work can still to be done. If necessary, one can need much finer meshes, at least on the small interfaces. To do it without spending too much time and/or memory, one can simply refine the meshes only on interfaces where it is required. It could also be useful to study the range of parameters allowing to have stable configurations. It would also be interesting to compare the energy of the clusters presented here and of ``more classical'' clusters with the same number of bubbles enclosing the same volumes. A compelling extension to the numerical examples would be to built more complicated clusters with several multiple bubbles. Finally, from a theoretical point of view, one can wonder if the existence of these clusters can be ``easily'' proven and if it is possible to have $n$-fold bubbles for any integer $n$.
\section{Acknowledgements}
I would like to thank Professor Frank Morgan for pointing out mistakes in the abstract and introduction.
\bibliography{genus_1_3_5_11}

@inbook{almgren_sullivan,
author = {Almgren, Fred and Sullivan, John},
title = {Visualization of soap bubble geometries},
year = {1993},
isbn = {026205048X},
publisher = {MIT Press},
address = {Cambridge, MA, USA},
booktitle = {The Visual Mind: Art and Mathematics},
pages = {79–83},
numpages = {5}
}

@article{almgren-taylor,
       author = {{Almgren}, Frederick J. and {Taylor}, Jean E.},
        title = "{The Geometry of Soap Films and Soap Bubbles}",
      journal = {Scientific American},
         year = 1976,
        month = jul,
       volume = {235},
       number = {1},
        pages = {82-93},
          doi = {10.1038/scientificamerican0776-82},
}

@article{brakke,
author = {Kenneth A. Brakke},
title = {{The surface evolver}},
volume = {1},
journal = {Experimental Mathematics},
number = {2},
publisher = {A K Peters, Ltd.},
pages = {141 -- 165},
year = {1992},
}

@book{cantat_cohen-addad_elias_graner_hoehler_pitois_rouyer_saint-jalmes_flatman_cox,
    author = {Cantat, Isabelle and Cohen-Addad, Sylvie and Elias, Florence and Graner, François and H\"{o}hler, Reinhard and Pitois, Olivier and Rouyer, Florence and Saint-Jalmes, Arnaud and Flatman, Ruth and Cox, Simon},
    title = {Foams: Structure and Dynamics},
    publisher = {Oxford University Press},
    year = {2013},
    month = {07},
    isbn = {9780199662890},
    doi = {10.1093/acprof:oso/9780199662890.001.0001},
    url = {https://doi.org/10.1093/acprof:oso/9780199662890.001.0001},
}

@article{hutchings-morgan-ritore-ros,
 ISSN = {0003486X},
 URL = {http://www.jstor.org/stable/3062123},
 author = {Michael Hutchings and Frank Morgan and Manuel Ritor\'{e} and Antonio Ros},
 journal = {Annals of Mathematics},
 number = {2},
 pages = {459--489},
 publisher = {Annals of Mathematics},
 title = {Proof of the Double Bubble Conjecture},
 urldate = {2024-12-09},
 volume = {155},
 year = {2002},
 doi = {10.2307/3062123},
}

@misc{milman-neeman,
      title={The Structure of Isoperimetric Bubbles on $\mathbb{R}^n$ and $\mathbb{S}^n$},
      author={Emanuel Milman and Joe Neeman},
      year={2024},
      eprint={2205.09102},
      archivePrefix={arXiv},
      primaryClass={math.DG},
      url={https://arxiv.org/abs/2205.09102},
}

@book{morgan-book,
  title={Geometric Measure Theory: A Beginner's Guide},
  author={Morgan, F.},
  isbn={978-0-12-804489-6},
  year={2016},
  publisher={Academic Press},
  doi={10.1016/C2015-0-01918-9},
}

@article{morgan-paper,
author = {Frank Morgan},
title = {{Immiscible fluid clusters in $\mathbb{R}^2$ and $\mathbb{R}^3$.}},
volume = {45},
journal = {Michigan Mathematical Journal},
number = {3},
publisher = {University of Michigan, Department of Mathematics},
pages = {441 -- 450},
year = {1998},
doi = {10.1307/mmj/1030132292},
URL = {https://doi.org/10.1307/mmj/1030132292}
}

@article{multitracker,
    author = "Fang Da and Christopher Batty and Eitan Grinspun",
    title = "Multimaterial Mesh-Based Surface Tracking",
    journal = {ACM Trans. on Graphics (SIGGRAPH North America 2014)},
    year = 2014
}

@inproceedings{plateau,
  title={Experimental and Theoretical Statics of Liquids Subject to Molecular Forces Only},
  author={Joseph Plateau},
  year={1873},
}

@misc{polyscope,
  title = {Polyscope},
  author = {Nicholas Sharp and others},
  note = {www.polyscope.run},
  year = {2019}
}

@article{schwarz,
author = {Schwarz, H. A.},
journal = {Nachrichten von der K\"{o}nigl. Gesellschaft der Wissenschaften und der Georg-Augusts-Universit\"{a}t zu G\"{o}ttingen},
pages = {1-13},
title = {Beweis des {S}atzes, dass die {K}ugel kleinere {O}berfl\"{a}che besitzt, als jeder andere {K}\"{o}rper gleichen {V}olumens},
url = {http://eudml.org/doc/180009},
volume = {1882},
year = {1882},
}

@inproceedings{sullivan,
  title={Nonspherical Bubble Clusters},
  author={John Sullivan},
  year={2014},
  url={https://api.semanticscholar.org/CorpusID:73713021}
}

@article{sullivan-morgan,
  title={OPEN PROBLEMS IN SOAP BUBBLE GEOMETRY},
  author={John M. Sullivan and Frank Morgan},
  journal={International Journal of Mathematics},
  year={1996},
  volume={07},
  pages={833-842},
  doi={10.1142/S0129167X9600044X},
}

@article{walters_davidson,
       author = {{Walters}, J.~K. and {Davidson}, J.~F.},
        title = "{The initial motion of a gas bubble formed in an inviscid liquid. Part 2. The three-dimensional bubble and the toroidal bubble}",
      journal = {Journal of Fluid Mechanics},
         year = 1963,
        month = jan,
       volume = {17},
        pages = {321-336},
          doi = {10.1017/S0022112063001373},
}

@book{weaire_hutzler,
  title={The Physics of Foams},
  author={Weaire, D.L. and Hutzler, S.},
  isbn={9780198510970},
  lccn={99033466},
  year={1999},
  publisher={Clarendon Press}
}
\end{document}